\documentclass[preprint,titlepage,aps,prd,amsmath,floats,floatfix,groupedaddress,superscriptaddress,nofootinbib,showpacs]{revtex4-2}

\usepackage{amssymb}
\usepackage{latexsym}
\usepackage[utf8]{inputenc}
\usepackage[english]{babel}
\usepackage{graphicx} % Include figure files
\usepackage{bm} % bold math
\usepackage{physics}
\usepackage{mathtools}
\usepackage{mathrsfs}
\usepackage{float}
\usepackage[style=base]{subcaption}

\usepackage[ruled,vlined]{algorithm2e}
\usepackage[left=2cm,right=2cm,top=2cm,bottom=2cm]{geometry}
\usepackage{setspace}

\usepackage{xcolor}
\definecolor{codegreen}{rgb}{0,0.6,0}
\definecolor{codegray}{rgb}{0.5,0.5,0.5}
\definecolor{codepurple}{rgb}{0.58,0,0.82}
\definecolor{backcolour}{rgb}{0.95,0.95,0.92}

\usepackage{listings}
\lstdefinestyle{mystyle}{
    backgroundcolor=\color{backcolour},   
    commentstyle=\color{codegreen},
    keywordstyle=\color{magenta},
    numberstyle=\tiny\color{codegray},
    stringstyle=\color{codepurple},
    basicstyle=\ttfamily\footnotesize,
    breakatwhitespace=false,         
    breaklines=true,                 
    captionpos=b,                    
    keepspaces=true,                 
    numbers=left,                    
    numbersep=5pt,                  
    showspaces=false,                
    showstringspaces=false,
    showtabs=false,                  
    tabsize=2
}
\usepackage{color}

\PassOptionsToPackage{hyphens}{url}\usepackage[breaklinks=true]{hyperref}
\hypersetup{pdfauthor={Santiago Ontanon and Miguel Alcubierre}}
\hypersetup{pdftitle={Rotating Boson Stars Using Finite Differences and Global Newton Methods}}

\numberwithin{equation}{section}
\numberwithin{figure}{section}
\numberwithin{table}{section}

\begin{document}

%%%%%%%%%%%%%%%%%%%%%%%%%%%%%%%%%
%%%   JOURNAL SPECIFICATION   %%%
%%%%%%%%%%%%%%%%%%%%%%%%%%%%%%%%%

% Classical and Quantum Gravity.
\preprint{IOP/CQG}

%%%%%%%%%%%%%%%%%
%%%   TITLE   %%%
%%%%%%%%%%%%%%%%%

\title{Rotating Boson Stars Using Finite Differences \texorpdfstring{\\}{\space}and Global Newton Methods}

\author{Santiago Ontañón}
\email{santiago.ontanon@correo.nucleares.unam.mx}

\author{Miguel Alcubierre}
\email{malcubi@nucleares.unam.mx}

\affiliation{Instituto de Ciencias Nucleares, Universidad Nacional
Aut\'onoma de M\'exico, A.P. 70-543, M\'exico D.F. 04510, M\'exico.}

%%%%%%%%%%%%%%%%
%%%   DATE   %%%
%%%%%%%%%%%%%%%%

\date{\today}

%%%%%%%%%%%%%%%%
%%%   PACS   %%%
%%%%%%%%%%%%%%%%


%%%%%%%%%%%%%%%%%%%%
%%%   ABSTRACT   %%%
%%%%%%%%%%%%%%%%%%%%

\begin{abstract}
We study Rotating Boson Star initial data for Numerical Relativity as previously considered by Yoshida and Eriguchi \cite{Yoshida_1997}, Lai \cite{Lai_2004}, and Grandclement, Somé and Gourgoulhon \cite{Grandclement_2014}. We use a 3 + 1 decomposition as presented by Gourgoulhon \cite{Gourgoulhon_2010} and Alcubierre \cite{Alcubierre_2012}, adapted to an axisymmetric quasi-isotropic spacetime with added regularization at the axis following work by Ruíz, Alcubierre and Núñez \cite{Ruiz_2008} and Torres \cite{Torres_2012}. The Einstein-Klein-Gordon equations result in a system of six coupled, elliptic, nonlinear equations with an added unknown for the scalar field's frequency $\omega$. Utilizing a Cartesian two-dimensional grid, finite differences, Global Newton Methods adapted from Deuflhard \cite{Deuflhard_2011}, the sparse direct linear solver PARDISO \cite{PARDISO_1,PARDISO_2}, and properly constraining all variables generates data sets for rotation azimuthal integers $l \in [0, 6]$. Our numerical implementation, published in GitHub \cite{ROTBOSON}, is shown to correctly converge both with respect to the resolution size and boundary extension (fourth order and third order, respectively). Thus, global parameters such as the Komar masses and angular momenta can be precisely calculated to characterize these spacetimes. Furthermore, analyzing the full family at fixed rotation integer produces maximum masses and minimum frequencies. These coincide with previous results in literature  for $l \in [0,2]$ as in \cite{Yoshida_1997,Lai_2004,Grandclement_2014,Liebling_2017} and are new for $l > 2$. In particular, the study of high-amplitude and localized scalar fields in axial symmetry is revealed to be only possible by adding the sixth regularization variable.
\end{abstract}

%%%%%%%%%%%%%%%%%%%%%%
%%%   MAKE TITLE   %%%
%%%%%%%%%%%%%%%%%%%%%%

\maketitle

%%%%%%%%%%%%%%%%%%%%%%%%
%%%   INTRODUCTION   %%%
%%%%%%%%%%%%%%%%%%%%%%%%

\section{Introduction}\label{sec:Introduction}

Boson stars are localized configurations of scalar fields first studied by Kaup \cite{Kaup_1968}, and Ruffini and Bonazzola \cite{Ruffini_1969}.  They arise as a natural coupling of a smooth classical complex scalar field with Einstein's theory of general relativity \cite{Einstein_1915}. Whereas Wheeler's original idea for a ``gravitational atom'', the \textit{geon} \cite{Wheeler_1955}, was proven to be unstable, for a scalar field the gravitational collapse can be avoided by the dispersive nature of the Klein-Gordon equation. Although Derrick's theorem \cite{Derrick_1964,Diez-Tejedor_2013} shows that no regular, static, and localized scalar fields are stable in three-dimensional space, boson stars solve this conundrum by demoting staticity to stationarity, and assuming a harmonic time behavior for the complex scalar field of the form
\begin{equation}
    \Phi(\vb{r},t) = \phi(\vb{r})\,e^{-i\omega t}\,,
\end{equation}
where $\omega$ is called the scalar field's frequency that arises as an eigenvalue when impossing the stationarity condition. Thus, boson stars are stationary and compact bundles of a complex scalar field, and given some hypothetical scalar field they may serve different astrophysical purposes, for example as mimics of neutron stars. Boson stars have also been used to model dark matter halos by Schunck and Torres~\cite{Schunck_2003} and Ureña-López and Bernal \cite{Urena-Lopez_2010}, and also as black-hole mimickers by Guzmán and Rueda-Becerril \cite{Guzman_2009} and Barranco and Bernal \cite{Barranco_2011}. On the question of their actual astrophysical existence, one must first consider the type of scalar field one has in mind. Indeed, for a free Klein-Gordon field a simple application of the uncertainty principle states that for a boson star with constituents of mass $m$ and localized within a Schwarzschild radius, the maximum mass must be of the order of 
\begin{equation}\label{eq:Kaup_limit}
    M_{\text{max}} \sim \frac{\hbar c}{2 G m} = \frac{m_P^2}{2m}\,,
\end{equation}
where $m_P \equiv \sqrt{\hbar c / m}$ is the Planck mass. For a scalar such as the Higgs boson, the resulting star mass is only $2\times 10^9$ kg, which is a tiny fraction of a Solar Mass. On the other hand, for an interacting field, such as the quartic interaction studied by Colpi \textit{et al.} \cite{Colpi_1986}, it has been shown that this limit can be increased almost arbitrarily to astronomically expected scales. However, throughout this paper we will only consider the case of a free Klein-Gordon field.

The simplest solution for a boson star is a \textit{spherical boson star}\/ that considers a spherically symmetric spacetime and scalar field such that $\phi(\vb{r}) = \phi(r)$. This problem is then one-dimensional and has been studied in the literature multiple times, including numerical results by Kaup \cite{Kaup_1968} and Friedberg \textit{et al.} \cite{Friedberg_1987}, using a shooting method that solves for an appropriate frequency such that the field and metric quantities decay correctly at spatial infinity. This solution is constrained by setting the field's value at the origin, $\phi_0 \equiv \phi(r=0)$, and solving the eigenvalue problem in order to find the frequency $\omega$, which results in a family of solutions parametrized by the value of $\phi_0$. These solutions have been shown to have a maximum mass of (as reported by Lai \cite{Lai_2004} and Grandclement \textit{et al.} \cite{Grandclement_2014})
\begin{equation}
    M_{\text{max}}^{l = 0} \approx 0.633 \: \frac{m_P^2}{m}\,,
\end{equation}
in agreement with the uncertainty principle limit mentioned above (equation \eqref{eq:Kaup_limit}).

Here we will consider the case of rotating boson stars which are a natural extension of the problem into an axisymmetric spacetime. These objects consider an additional harmonic decomposition in the rotation angle $\varphi$,
\begin{equation}
    \Phi(\vb{r},\,t) = \phi(\vb{r})\,e^{-i(\omega t - l \varphi)}\,,
\end{equation}
where $l$ is an integer known as the azimuthal rotation number. As expected, rotating boson stars have non-zero angular momentum and greater complexities than their spherical siblings (notice that this is quite different from the $\ell$-boson stars recently studied by Alcubierre \textit{et al.} \cite{Alcubierre_2018}, which are particular combinations of several scalar fields that result in spherical objects with non-zero total angular momentum).

Rotating boson stars were first studied by Silveira and Sousa in Newtonian theory \cite{Silveira_1995}. The first general relativity numerical results were obtained by Yoshida and Eriguchi for the $l = 1$ case \cite{Yoshida_1997}, for which they found a maximum mass of $M_{\text{max}}^{l=1} \approx 1.31\,m_P^2/m$. Further analysis was done by Lai in his PhD thesis \cite{Lai_2004} where he extended results to the $l = 2$ case, and also found a maximum mass for $l = 1$, although it differed from the result obtained by Yoshida and Eriguchi. Mielke and Schunck \cite{Schunck_2003} found solutions all throughout $l \in [1,10]$ (and also for $l = 500$), however their results were limited to extremely small amplitude near-Newtonian cases. Finally, Grandclement, Somé and Gourgoulhon \cite{Grandclement_2014} found rotating boson stars for $l \in [1,4]$ which confirmed Yoshida and Eriguchi's $l = 1$ results, and also estimated the maximum mass for $l = 2$. More background and information on boson stars can be found in Liebling and Palenzuela's review \cite{Liebling_2017}, which covers the widest breadth of topics (whereas Mielke's review considers rotating boson stars exclusively \cite{Mielke_2016}).

For this paper we have considered free-field boson stars with rotation azimuthal number $l \in [1,6]$. Differing from Lai's and Grandclement \textit{et al.}'s approach, we have chosen to use a non-compactified spacetime where the grid is a simple uniform grid with finite difference discretization. However, in reading both aforementioned works, the issue of the field's regularity at the axis was posited as a reason for numerical failure at high amplitudes. Thus, by looking into previous works by Ruíz, Alcubierre, and Núñez \cite{Ruiz_2008}, the regularization of the axis has been a major component in our analysis.

This article is the organized as follows. In Section \ref{sec:Field Equations} we present the field equations and the deduction of the system of five nonlinear elliptic equations that describe rotating boson stars.  Section \ref{sec:Boundary Conditions} is key in presenting our boundary conditions, since an inadequate constraint will lead to trivial Minkowski spacetime or excited boson stars states. Section \ref{sec:Regularization} introduces a sixth variable that guarantees regularity at the axis even for high amplitudes, and will prove critical for extending previous results. Section \ref{sec:Global Quantities}
details how masses and angular momenta were calculated given our grid limitations. Section \ref{sec:Global Newton Methods} presents our numerical algorithm for obtaining initial data for a highly nonlinear problem where initial guesses are hard to construct. Having presented both the theory and numerical algorithms, Section \ref{sec:Code Overview} gives and explanation and overview of the numerical code, \textit{ROTBSON} specially written for rotating boson star initial data generation. Section \ref{sec:Error Indicators} presents \textit{ROTBOSON}'s expected fourth-order convergence and gives estimates for relative errors in global quantities. Finally, our results are presented in Section \ref{sec:Results}, and Section \ref{sec:Summary and Conclusions} gives some concluding remarks on these results.

A final comment on conventions and units. Throughout this work we use geometrized units where $c = G = \hbar = 1$, unless stated otherwise. In these units, the Planck mass is given by $m_{P} = \sqrt{\hbar c / G} = 1$. This implies that our coordinate lengths will have dimensions directly inverse to the field's mass, $(1/m)$. In the results section we recover Planck's reduced constant to present them as is usual in the literature, such as in \cite{Grandclement_2014}. Lastly, although all calculations are invariant to this, our metric signature is the usual convention for general relativity $(-,\,+,\,+,\,+)$.

%%%%%%%%%%%%%%%%%%%%%%%%%%
%%%   FIELD EQUATIONS  %%%
%%%%%%%%%%%%%%%%%%%%%%%%%%

\section{Field Equations}\label{sec:Field Equations}

We use the 3+1 general relativity formalism as presented by York \cite{York_1978}, Alcubierre \cite{Alcubierre_2012}, and Gourgoulhon \cite{Gourgoulhon_2010}, where the spacetime metric is written as
\begin{equation}\label{eq:spacetime_metric}
\dd s^2 = g_{\mu \nu}\,\dd x^\mu\,\dd x^\nu = (-\alpha^2 + \gamma_{i j}\,\beta^i\,\beta^j)\,\dd t^2 + 2\gamma_{ij}\,\beta^j\,\dd t\,\dd x^i + \gamma_{ij}\,\dd x^i\,\dd x^j\,.
\end{equation}
Here $\alpha$ is the lapse function, $\beta^i$ is the shift vector, and $\gamma_{ij}$ is the spatial metric induced on hypersurfaces of constant coordinate time $\Sigma_t$. Furthermore, we consider a stationary and axisymmetric spacetime. In an adapted coordinate system, we can generate these symmetries with the stationary generator $(\pdv*{t})^\mu = \alpha\,n^\mu + \beta^\mu$ (where $n^\mu$ is the normal vector to the spatial hypersurfaces $\Sigma_t$) and the axisymmetric generator $(\pdv*{\varphi})^\mu$. An extensive study of these symmetries and formalism can be found in \cite{Gourgoulhon_2010}. From this same reference, we use quasi-isotropic coordinates $(t,\,\rho,\,z,\,\varphi)$, also called Lewis-Papapetrou coordinates \cite{Lewis_1932,Papapetrou_1945}. Notice that we are using cylindrical coordinates $(\rho,\,z)$ instead of polar $(r,\,\theta)$. However, this is perfectly equivalent with the usual transformation $\rho = r\,\sin\theta$ and $z = r\,\cos\theta$. In this adapted coordinate system the shift vector can be written as
\begin{equation}\label{eq:shift}
\beta^i = \Omega(\rho,\,z)\,(\pdv*{\varphi})^i\,,
\end{equation}
whereas the spatial metric takes the simple form
\begin{equation}\label{eq:spatial_metric}
\dd l^2 = \gamma_{ij}\,\dd x^i\,\dd x^j \\= A(\rho,\,z)\,(\dd \rho^2 + \dd z^2) + \rho^2\,H(\rho,\,z)\,\dd \varphi^2\,.
\end{equation}

Thus, the spacetime geometry is reduced to four functions $(\alpha,\,\Omega,\,A,\,H)$ of the two coordinates $(\rho,\,z)$. This is not all, since the Einstein equation will couple geometry to matter via the stress-energy tensor $T_{\mu \nu}$. Our scalar field will be written as the standard ansatz with harmonic time and angular dependence \cite{Lai_2004,Mielke_2016}
\begin{equation}\label{eq:scalar_field}
\Phi(t,\,\rho,\,z,\,\varphi) = \phi(\rho,\,z)\,e^{-i(\omega t - l \varphi)}\,.
\end{equation}
Above, $l$ must be an integer for single-valuedness at $\varphi = 0,\,2\pi$, and is known as the azimuthal rotational number (sometimes referred to as $k$ or $m$ in other works).

\smallskip

Thus, we need five equations for the five unknown variables $\lbrace \alpha,\,\Omega,\,A,\,H,\,\phi \rbrace$. These are given via:

\begin{enumerate}
\item The Lewis-Papapetrou or quasi-isotropic coordinates imply that $K$, the trace of the extrinsic curvature \mbox{$K_{ij} = -\tfrac{1}{2}\,\pounds_n \gamma_{ij}$}, is zero, which results in the maximal slicing condition,
\begin{equation}\label{eq:maximal_slicing}
D^2\,\alpha - \alpha\,\left(4\pi\,(u + S) + K_{ij} K^{ij}\right) = 0\,,
\end{equation}

\item The Hamiltonian constraint, 
\begin{equation}\label{eq:hamiltonian_constraint}
\mathcal{H} = {}^{(3)}R + K^2 - K_{ij} - 16\pi\,u = 0\,,
\end{equation}

\item The momentum constraint for $i = \varphi$, 
\begin{equation}\label{eq:momentum_constraint}
\mathcal{M}^i = D_j\,K^{ij} - D^i\,K - 8\pi\,j^i = 0\,,
\end{equation}

\item The stationarity condition $\partial_t K_{ij} = 0$ for $i = j = \varphi$,
\begin{equation}\label{eq:K_time_derivative}
\pdv{K_{ij}}{t} = \pounds_\beta K_{ij} - D_i D_j \,\alpha + \alpha\,\left(^{(3)}R_{ij} + K\,K_{ij} - 2K_{ik}\,K^k\,_j + 4\pi\,\left((S - u)\,\gamma_{ij} - 2 S_{ij}\right)\right) = 0\,,
\end{equation}

\item The Klein-Gordon equation
\begin{equation}\label{eq:klein-gordon}
\left( \nabla^2 - m^2\right)\,\Phi = 0\,.
\end{equation}

\end{enumerate}

\noindent Above, $D_i$ is the covariant derivative associated to the spatial metric $\gamma_{ij}$, $^{(3)}R_{ij}$ its Ricci tensor, and $u,\,j^i,\,S_{ij}$ are respectively the energy density, momentum density and stress tensor as seen by the Eulerian observers, defined as the following projections of $T_{\mu \nu}$ \cite{Alcubierre_2012}
\begin{align}\label{eq:zamo_decomposition}
u \;=&\, n^\mu n^\nu\,T_{\mu \nu}\,,\\
j^i \;=&\, -\gamma^{i \mu}\,n^\nu\,T_{\mu \nu}\,,\\
S_{ij} =&\, \gamma^\mu\,_i\,\gamma^\nu\,_j\,T_{\mu \nu}\,,\\
S \;=&\, \gamma^{ij}\,S_{ij}\,,
\end{align}
and $m$ is the scalar field's mass parameter.

\smallskip

To obtain the final equations as presented below, linear combinations of these five equations must be taken. The resulting system of equations is
\begin{align}
\begin{split}\label{eq:f_alpha}
f_\alpha \equiv&\, \left(\pdv[2]{\rho} + \pdv[2]{z} + \frac{1}{\rho}\,\pdv{\rho} \right)\,\alpha + \frac{1}{2H}\,\partial \alpha\,\cdot\,\partial H - \frac{\rho^2 H}{2\alpha}\,\partial\Omega\,\cdot\,\partial \Omega\\
&  - 4\pi\,A\,\left(\frac{2(\omega + l\Omega)^2}{\alpha} - m^2\,\alpha\right)\,\phi^2 = 0\,,
\end{split}\\
\begin{split}\label{eq:f_Omega}
f_\Omega \equiv&\, \left(\pdv[2]{\rho} + \pdv[2]{z} + \frac{3}{\rho}\,\pdv{\rho} \right)\,\Omega+ \frac{3}{2H}\,\partial \Omega\,\cdot\,\partial H - \frac{1}{\alpha}\,\partial \Omega\,\cdot\,\partial \alpha - 16\pi\,\frac{A}{H}\,l\,(\omega + l\Omega)\,\left(\frac{\phi}{\rho}\right)^2 = 0\,,
\end{split}\\
\begin{split}\label{eq:f_A}
f_A \equiv&\, \left(\pdv[2]{\rho} + \pdv[2]{z}\right)\,A - \frac{1}{A}\,\partial A\,\cdot\,\partial A + \left[ -\frac{1}{\alpha H}\,\partial H\,\cdot\,\partial \alpha - \frac{\rho^2 H}{2\alpha^2}\,\partial\Omega\,\cdot\,\partial \Omega - \frac{2}{\rho\, \alpha}\,\partial_\rho \alpha\right.\\
&\left. + 8\pi\,\left\lbrace \left(\frac{\phi}{\rho}\right)^2\,\left(\rho^2\,\frac{(\omega + l \Omega)^2}{\alpha^2}\,A - l^2\,\frac{H}{A}\right) + \partial\phi\,\cdot\,\partial\phi\right\rbrace \right]\,A = 0\,,
\end{split}\\
\begin{split}\label{eq:f_H}
f_H \equiv&\, \left(\pdv[2]{\rho} + \pdv[2]{z} + \frac{2}{\rho}\,\pdv{\rho} \right)\,H - \frac{1}{2H}\,\partial H\,\cdot\,\partial H + \frac{1}{\alpha}\,\partial H\,\cdot\,\partial \alpha \\
&\quad + \left(\frac{\rho^2 H}{\alpha^2}\,\partial \Omega\,\cdot\,\partial \Omega + \frac{2}{\rho\,\alpha}\,\partial_\rho \alpha + 8\pi\,A\,\left(\frac{\phi}{\rho}\right)^2\left(\rho^2m^2 + \frac{2l^2}{H}\right)\right)\,H = 0\,,
\end{split}\\
\begin{split}\label{eq:f_phi}
f_\phi \equiv&\, \left(\pdv[2]{\rho} + \pdv[2]{z} + \frac{1}{\rho}\,\pdv{\rho} - \frac{l^2}{\rho^2}\right)\,\phi  + \frac{1}{\alpha}\,\partial \alpha\,\cdot\,\partial \phi + \frac{1}{2H}\,\partial H\,\cdot\,\partial \phi\\
& + \left(A\,\left(\frac{(\omega + l \Omega)^2}{\alpha^2} - m^2\right) - \frac{l^2}{H}\,\left(\frac{A - H}{\rho^2}\right)\right)\,\phi = 0\,.
\end{split}
\end{align}
These equations use the short-hand defined in \cite{Grandclement_2014,Gourgoulhon_2010}:
\begin{equation}\label{eq:pdv_product}
\partial u\,\cdot \,\partial v \equiv \pdv{u}{\rho}\,\pdv{v}{\rho} + \pdv{u}{z}\,\pdv{v}{z}\,.
\end{equation}

Equations \eqref{eq:f_alpha} - \eqref{eq:f_phi} are not quite the final equations we will be using. In fact, $\lbrace \alpha,\, \Omega,\, A,\, H,\,\phi \rbrace$ are also not the actual variables used in our work. Instead we will work with
\begin{align}\label{eq:real_variables}
\aleph \equiv&\, \log \alpha\,,\\
a \equiv&\, \frac{1}{2}\,\log A\,,\\
h \equiv&\, \frac{1}{2}\,\log H\,,\\
\psi \equiv&\, \frac{\phi}{\rho^l}\,.
\end{align}

The first three variables are introduced to guarantee that $\alpha,\, H,\,A > 0$ (as should be from metric-positivity). The final variable comes from examining regularity conditions in Eq.~\eqref{eq:f_phi}. More precisely, the first four terms that make up the elliptic operator in this equation imply that, if everything else in the equation is regular at the rotation axis $\rho = 0$, $\phi$ must then vanish as $\rho^l$ on this same axis. This might not seem true, considering the term $(A - H)/\rho^2$, however, this term is shown to be regular at the axis via local-flatness, as will be further discussed in Section \ref{sec:Regularization} below. The immediate implication of this decomposition is that the term $\phi / \rho = \rho^{l - 1}\,\psi$ is regular for $l \geq 1$, which is true for all rotating boson stars. Using this new $\psi$ variable, its resulting equation is
\begin{align}\label{eq:f_psi}
\begin{split}
f_\psi \equiv&\, \left(\pdv[2]{\rho} + \pdv[2]{z} + \frac{(2l + 1)}{\rho}\,\pdv{\rho} \right)\,\psi + \frac{1}{\alpha}\,\partial \alpha\,\cdot\,\partial \psi + \frac{1}{2H}\,\partial H\,\cdot\,\partial \psi\\
& + \left(\frac{l}{\rho}\,\left(\frac{\partial_\rho \alpha}{\alpha} + \frac{\partial_\rho H}{2H}\right) + A\,\left(\frac{(\omega + l \Omega)^2}{\alpha^2} - m^2\right) - \frac{l^2}{H}\,\left(\frac{A - H}{\rho^2}\right)\right)\,\psi = 0\,.
\end{split}
\end{align}
Notice also that using $\psi$ is convenient since otherwise the field's first $l$ derivatives on the $\rho$ direction would need to vanish, something very difficult to guarantee numerically. This was reported to be an issue in Yoshida and Eriguchi's first work \cite{Yoshida_1997}.

We will not rewrite equations \eqref{eq:f_alpha} - \eqref{eq:f_H} and \eqref{eq:f_psi} in terms of $\lbrace \aleph,\,\Omega,\,a,\,h,\,\psi \rbrace$ since the change of variables is quite simple. Thus, we have a set of five nonlinear elliptic equations to solve for the initial data. Unfortunately, this system has a trivial solution given by $\alpha = A = H = 1$, $\Omega = \phi = 0$, which corresponds to Minkowski spacetime. We obviously do not wish to obtain this spacetime configuration, so a first step is to constrain solutions to have $|\phi| > 0$, or, equivalently $|\psi| > 0$. In fact, we can be more specific, since we are searching for ground-state solutions where the scalar field has no ``nodes'' (save at the rotation axis) and thus we will ask for $\psi > 0$. This constraint will be further developed in the next section.

%%%%%%%%%%%%%%%%%%%%%%%%%%%%%%
%%%   BOUNDARY CONDITIONS  %%%
%%%%%%%%%%%%%%%%%%%%%%%%%%%%%%

\section{Boundary Conditions}\label{sec:Boundary Conditions}

First a comment on our grid structure: we will be solving this system of equations with a two-dimensional cartesian $(\rho,\,z)$ uniform grid. We could use a more complex non-uniform or adaptive grid, but for simplicity's sake, we have opted to use the easier approach where the grid spaces are constant and given by $\Delta \rho$ and $\Delta z$ in their respective directions. As is common for Numerical Relativity, a staggered grid is used with the goal of avoiding possible problematic divisions by zero on the $\rho$ axis, and also to allow us to impose equatorial symmetry in a simple way. Thus, our discrete coordinates are given by 
\begin{equation}\label{eq:discrete_coordinates}
(\rho_i,\,z_j) = \left(\Delta \rho\,(i - g + \tfrac{1}{2}),\,\Delta z\,(j - g + \tfrac{1}{2})\right)\,,
\end{equation}
where $g$ is the number of ghost zones necessary for finite differences, and $i$, $j$ are integers in the intervals:
\begin{equation}\label{eq:indices_range}
i\in [0,\,N_\rho + 2g - 1]\,, \quad j \in [0,\,N_z + 2g - 1]\,.
\end{equation}
This means that $N_\rho,\,N_z$ are the number of interior points in each respective direction, and we have two ghost zones boundary bands with the first interior point being $(i,\,j) = (g,\,g)$ and the last interior point corresponding to $(i,\,j) = (g+N_\rho-1,\,g+N_z-1)$.

\smallskip

Notice that our system of equations also has equatorial symmetry. Thus, the most optimal solution algorithm only solves for a single quadrant of the $(\rho,\,z)$ coordinates, namely, the positive quadrant. The remaining quadrants can be obtained by parity transformations. It is easy to see that invariance under the independent symmetries $\rho\to -\rho$ and $z \to -z$ implies that all our functions are even functions of both variables. This is a boundary condition for our grid, specifically for our left and bottom strips. The mathematical expression is
\begin{equation}\label{eq:parity_condition}
u(i,\,j) = u(2g - 1 - i,\,j)\,,\, i\in [0,\,g - 1]\,,\quad u(i,\,j) = u(i,\,2g - 1 - j)\,,\, j\in [0,\,g -1]\,,
\end{equation}
for $u \in \lbrace \aleph,\,\Omega,\,a,\,h,\,\psi \rbrace$. 

\smallskip

We also need boundary conditions for the external boundaries $i \geq N_\rho + g,\, j \geq N_z + g$. Asymptotic flatness has the simple condition $\alpha,\, A,\, H = 1$ and $\Omega = \psi = 0$ at spatial infinity. However, we do not have a compactified spacetime and must use another condition instead. Enforcing the previous condition as a Dirichlet approximation at finite distance is not ideal, rather we use the Robin boundary conditions \cite{Gustafson_1998,Nora_2018} that come from expanding a grid variable $u$ asymptotically for large $r$ as:
\begin{equation}\label{eq:robin_series}
u(r,\,\theta) = u_\infty + \frac{C_n}{r^n} + \mathcal{O}(r^{-(n + 1)},\,\theta)\,,
\end{equation}
where $u_\infty$ is the value or limit of $u$ at spatial infinity, $C_n$ is a constant independent of the direction $\theta$, $n$ a positive integer, and further terms can depend on the direction $\theta$ but are of order $r^{-(n+1)}$ or greater.

\smallskip

The actual Robin boundary condition is written as a mix of Dirichlet and Neumann conditions which is a natural consequence of the above Equation \eqref{eq:robin_series}:
\begin{equation}\label{eq:robin_bc}
r_{\text{bdy}}\,\pdv{u}{r}\,(r_{\text{bdy}},\,\theta) + n\,(u(r_{\text{bdy}},\,\theta) - u_\infty) = \mathcal{O}(r_{\text{bdy}}^{-(n + 1)},\,\theta)\,.
\end{equation}
Above, $r_{\text{bdy}}$ stands for an external boundary point such that $r_{\text{bdy}} \gg 1$. The numerical implementation sets the right-hand side of \eqref{eq:robin_bc} to zero and, for our two dimensional grid, we have $r\,\pdv*{u}{r} = \rho\,\pdv*{u}{\rho} + z\,\pdv*{u}{z}$.

\smallskip

Notice that the unknown constant $C_n$ has dropped out from our boundary condition. This is not so for the integer $n$, which must be known \textit{a priori}. This integer can be calculated from expanding the system of equations around spatial infinity and determining the first nonzero power of $1/r$ for each variable. This is tedious since we first have to rewrite the equations in $(r,\,\theta)$ coordinates, but quite straightforward. In this way we find that $n = 1$ for $\aleph,\,a,\,h$, and $n = 3$ for $\Omega$. In fact, it can also be shown that the constant $C_n$ must have the same value for $a$ and $h$, and minus this value for $\aleph$. This will be no surprise when we talk about global quantities in Section \ref{sec:Global Quantities}. Also, the Dirichlet condition at spatial infinity means that all variables $\lbrace \aleph,\,\Omega,\,a,\,h\rbrace$ have $u_\infty = 0$.

\smallskip

We have yet to deal with the scalar field. Its decay is in fact not a power of $1/r$, but instead exponential. This can be seen by expanding equation \eqref{eq:f_phi} at spatial infinity and keeping only the lowest order in $1/r$, i.e. $\alpha = A = H = 1$ and $\Omega = 0$. This leads to 
\begin{equation}\label{eq:f_phi_asymptotic}
\left(\pdv[2]{r} + \frac{2}{r}\,\pdv{r} + \frac{1}{r^2}\,\pdv[2]{\theta} + \frac{1}{r^2\,\tan\theta}\,\pdv{\theta} - \frac{l^2}{r^2\sin^2\theta} - (m^2 - \omega^2)\right)\,\phi \approx 0\,.
\end{equation}
We recognize the Laplacian operator in cylindrical coordinates above. By separation of variables, one can show that the appropriate solution of \eqref{eq:f_phi_asymptotic} is
\begin{equation}\label{eq:phi_asymptotic_solution}
\phi = C\,P^l_l(\cos\theta)\,k_{l}(\sqrt{m^2-\omega^2}\,r)\,,
\end{equation}
where $P^l_l(\cos\theta)$ is the associated Legendre polynomial, $k_{l}(\sqrt{m^2-\omega^2}\,r)$ is the modified spherical Bessel function, and $C$ is a constant. This solution is the appropriate one since it goes to zero at spatial infinity, and vanishes as $\sin^l \theta$ on the rotation axis. Indeed, from \cite{Arfken_2005}, given the asymptotic form of the Bessel function $k_l(x) \sim e^{-x}/x$, and a direct calculation of $P^l_l(\cos\theta) \sim \sin^l \theta$, one finds,
\begin{equation}\label{eq:phi_asymptotic_behavior}
\phi \to C\,\sin^l \theta\,\frac{e^{-\sqrt{m^2 - \omega^2}\,r}}{r}\,,
\end{equation}
or, in terms of our variable $\psi = \rho^{-l}\,\phi$,
\begin{equation}\label{eq:psi_asymptotic_behavior}
\psi \to C\,\frac{e^{-\sqrt{m^2 - \omega^2}\,r}}{r^{l+1}}\,.
\end{equation}
Observe that this asymptotic behavior requires that (reinserting Planck's Constant)
\begin{equation}\label{eq:omega_limits}
0 < \omega < \frac{m}{\hbar}\,,
\end{equation}
as otherwise there would not be an appropriate exponential decay of the scalar field. Thus, the differential equation for the boundary condition for $\psi$ is
\begin{equation}\label{eq:exp_decay_bc}
r_{\text{bdy}}\,\pdv{\psi}{r}\,(r_{\text{bdy}},\,\theta) + \left((l + 1) + r_{\text{bdy}}\,\sqrt{m^2 - \omega^2}\right)\,\psi(r_{\text{bdy}},\,\theta) = \mathcal{O}(r_{\text{bdy}}^{-(l+2)},\,\theta)\,.
\end{equation}

\smallskip

In conclusion, the above boundary conditions have given us a full list of the constraints our variables must satisfy:

\begin{enumerate}
\item The metric requires $\alpha,\,A,\,H$ positive, and decaying as $1/r$ asymptotically.

\item Gourgoulhon \cite{Gourgoulhon_2010} shows that in terms of our notation we must have $\Omega < 0$. Also, $\Omega$ decays as $1/r^3$.

\item The scalar field auxiliary variable $\psi$ must always be positive to guarantee that we have the ground-state, and must decay as $e^{-\sqrt{m^2-\omega^2}\,r}/r^{l+1}$. Nevertheless, to avoid the trivial Minkowski solution, $\psi$ must still be nonzero at the boundary, however small.

\item Finally, the field's frequency is bounded by the scalar field mass parameter.
\end{enumerate}

Condition 1 is satisfied by taking logarithmic variables as previously mentioned. Conditions 2 and 3 could be guaranteed by taking another variable such as $\log(-\Omega)$ and $\log(\psi\,e^{\sqrt{m^2-\omega^2}\,r})$. However, this leads to Robin boundary conditions where the value at spatial infinity is infinite. Therefore, we do not constrain these variables in practice, but violations of the aforementioned constraints would show that we are solving the wrong problem. Lastly, Condition 4 is affirmed by working with a new sigmoid variable that guarantees it
\begin{equation}\label{eq:xi}
\xi \equiv \tanh^{-1}\,\left(2\,\frac{\omega}{m} - 1\right)\, \quad\Rightarrow\quad \omega = \frac{m}{2}\,\left(1 + \tanh(\xi)\right)\,.
\end{equation}

%%%%%%%%%%%%%%%%%%%%%%%%%%
%%%   REGULARIZATION   %%%
%%%%%%%%%%%%%%%%%%%%%%%%%%

\section{Regularization}\label{sec:Regularization}

Before proceeding with technical and algorithmic details, we should carefully examine that all terms in our equations are not only analytically regular at both axes, but also, that we can write them in a fashion that is less likely to cause numerical problems. Terms such as $(1/\rho)\,\pdv*{u}{\rho}$ (for a given $u \in \lbrace \aleph,\, \Omega,\,a,\,h,\psi \rbrace$) are regular since all grid functions are even about the rotation axis. Therefore, $\pdv*{u}{\rho}$ behaves as $\mathcal{O}(\rho)$ at this same axis, and its combined division by $\rho$ is an even function. Nevertheless, it is recommended to write these terms with the order of operations explicit in numerical code, i.e., instead of writing in source code
\begin{equation*}
\frac{l}{\rho}\,\left(\pdv{\aleph}{\rho} + \pdv{h}{\rho}\right)\,\psi\,,
\end{equation*}
\noindent write
\begin{equation*}
l\,\left(\left(\frac{1}{\rho}\,\pdv{\aleph}{\rho}\right) + \left(\frac{1}{\rho}\,\pdv{h}{\rho}\right)\right)\,\psi\,.
\end{equation*}

As previously stated, terms such as $\phi/\rho$ are regular since the scalar field vanishes at least as $\rho$ on the axis. With this in mind, it is easy to verify that all terms in our equations are regular, with the possible exception of a single term in Equation \eqref{eq:f_psi} proportional to:
\begin{equation}\label{eq:lambda}
\lambda \equiv \frac{A - H}{\rho^2}\,.
\end{equation}

An initial approximation might seem to indicate that this should not be an issue, given that we are using a staggered grid. Thus, no actual division by zero is ever done. Indeed, the smallest value of the $\rho$ coordinate is, from \eqref{eq:discrete_coordinates}, $\rho = \rho_g = \Delta \rho/2$. Thus, we should not see regularization problems for low resolutions, i.e., for $\Delta \rho$ relatively big (empirically, $\Delta \rho \gtrsim 0.1$). However, as almost all problems in this field, a greater resolution is best to reduce truncation errors, and it is also necessary to closely examine small regions of spacetime, specially as the scalar field amplitude is increased.

Furthermore, as will be seen in Section \ref{sec:Results}, as we consider scalar fields with greater amplitude, the field concentrates closer to the rotation axis and develops a sharp maximum, i.e., the $\rho$ coordinate location of the field's maximum becomes smaller and smaller. Thus, we need greater resolution to properly resolve and characterize this spike. As we increase the resolution, $\Delta \rho$ decreases and we start seeing regularization problems in equation \eqref{eq:f_psi} coming from the term \eqref{eq:lambda}.

It is worthwhile to point out that these regularization issues come from our curvilinear coordinates. In a full 3D Cartesian scheme, $1/\rho$ or $1/r$ terms are absent. However, curvilinear coordinates have an obvious advantage when dealing with symmetric spacetimes such as axisymmetry where the Killing vectors are fully adapted. This means that a 3D calculation is reduced to 2D (for our axysimmetric work), making it far less computationally expensive.

For an exact solution the term \eqref{eq:lambda} is in fact perfectly regular.  Indeed, a simple argument of local flatness shows that near the rotation axis, we can expand $A$ and $H$ as
\begin{equation}\label{eq:axis_cancellation}
A(\rho,\,z) = \gamma_{\rho \rho} = C(0,\,z) + \frac{\rho^2}{2}\,A_2(0,\,z) + \mathcal{O}(\rho^4)\,,\quad H(\rho,\,z) = \frac{\gamma_{\varphi \varphi}}{\rho^2} = C(0,\,z) + \frac{\rho^2}{2}\,H_2(0,\,z) + \mathcal{O}(\rho^4)\,,
\end{equation}
i.e., from local flatness the value at the rotation axis, $C(0,\,z)$ must be equal for both metric functions, and thus $\lambda$ is a regular and even function at the rotation axis,
\begin{equation}\label{eq:lambda_expansion}
\lambda(\rho,\,z) = \frac{(A_2(0,\,z) - H_2(0,\,z))}{2} + \frac{\rho^2}{2}\,\lambda_2(0,\,z) + \mathcal{O}(\rho^4)\,.
\end{equation}

This clearly implies that the term in \eqref{eq:f_psi}, $-(l^2\,\lambda/H)\,\psi$, is \textit{analytically} regular at the axis. But, unfortunately, numerical methods introduce truncation and round-off errors that may cause that this exact cancellation will not hold numerically, and an increase in resolution will introduce loss of significance if we divide $(A-H)$ by a small quantity. To deal with this potential catastrophe, we follow previous work by Milton, Alcubierre and Núñez \cite{Ruiz_2008}, and Torres \cite{Torres_2012}, whence the quantity $\lambda$ is promoted to a new independent variable. This is expensive in terms of computational resources since we are introducing, intuitively, at least $20\%$ more work. Also, $\lambda$ must have its own elliptic equation and boundary conditions, otherwise the problem is indeterminate. Obtaining this equation is tedious work and nontrivial. A first guess is to take equations $f_A$ and $f_H$, \eqref{eq:f_A} and \eqref{eq:f_H}, subtract them and divide them by $\rho^2$. Afterwards, we construct an elliptic operator for $\lambda$, i.e., $\pdv*[2]{\lambda}{\rho}+ \pdv*[2]{\lambda}{z}$, and regularize all terms. For example, terms that initially appear irregular can be combined into a regular expressions,
\begin{equation}\label{eq:irregular_to_regular_combination}
\frac{1}{\rho^2}\,\left(\pdv[2]{\aleph}{\rho} - \frac{1}{\rho}\,\pdv{\aleph}{\rho}\right) = \frac{1}{\rho}\,\pdv{\rho}\,\left(\frac{1}{\rho}\,\pdv{\aleph}{\rho}\right)\,.
\end{equation}
Above we have the derivative of an even function, since we previously noted that $(1/\rho)\,\pdv*{\aleph}{\rho}$ was regular and even. Therefore, its derivative is odd, and its division by $\rho$ turns out to be an overall regular expression.

Unfortunately, the combination $(f_A - f_H) / \rho^2$ turns out to be inappropriate to reduce via this procedure in order to generate a regular equation. Happily, however, we have a large amount of equations to work with. Previously we specified the five equations that give rise to our system of equations. But there are in fact many more equations. Some are trivial, such as the momentum constraints $M^i$ for $i = \rho,\,z$, but stationarity $\partial_t K_{ij} = 0$ yields more equations for $i,j \neq \varphi$. The proper combination to obtain a regular equation for $\lambda$ turns out to be:
\begin{equation}\label{eq:f_lambda_short}
f_\lambda \equiv - \frac{2A}{\rho^2 \alpha}\,\left( \partial_t K_{\rho \rho} - \frac{1}{\rho^2}\,\partial_t K_{\varphi \varphi}\right) = \pdv[2]{\lambda}{\rho} + \pdv[2]{\lambda}{z} + \frac{3}{\rho}\,\pdv{\lambda}{\rho} + \ldots = 0\,.
\end{equation}
This is a rather large equation so we will not write it explicitly here (the full equation is given in Appendix \ref{sec:Appendix 1}, Equation \eqref{eq:f_lambda}). For our discussion, we only want to notice the elliptic operator in Equation \eqref{eq:f_lambda_short}.

Before continuing, we must specify boundary conditions for $\lambda$. We have already determined that $\lambda$ is an even function of $\rho$. By examining the full equation \eqref{eq:f_lambda}, we can also deduce that it is even about the $z$ axis, which is fortunate since otherwise we would need to solve another quadrant of $(\rho,\,z)$ space. For its external boundary condition, now that we have the asymptotic behavior of the other variables, a similar expansion at spatial infinity shows that $\lambda$ must decay as $1/r^4$, i.e. 
\begin{equation}\label{eq:lambda_robin_bc}
r_{\text{bdy}}\,\pdv{\lambda}{r}\,(r_{\text{bdy}},\,\theta) + 4\lambda(r_{\text{bdy}},\,\theta) = \mathcal{O}(r_{\text{bdy}}^{-5},\,\theta)\,.
\end{equation}

%%%%%%%%%%%%%%%%%%%%%%%%%%%%%
%%%   GLOBAL QUANTITIES   %%%
%%%%%%%%%%%%%%%%%%%%%%%%%%%%%

\section{Global Quantities}\label{sec:Global Quantities}

It is of great interest to characterize our solutions with a set of parameters that globally describe the spacetime. This can be done via conserved quantities which for General Relativity are associated with spacetime symmetries. Possibly the most general concept is the Arnowitt-Deser-Misner (ADM) mass  which only requires that the spacetime be asymptotically flat. From Wald \cite{Wald_1984}, this mass is defined as
\begin{equation}\label{eq:ADM_mass}
M_{\text{ADM}} = \lim_{\mathscr{S}\to \infty}\,\frac{1}{16\pi}\,\oint_{\mathscr{S}}\,\left(\mathcal{D}^j\,\gamma_{ij} - \mathcal{D}_i\,(f^{kl}\gamma_{kl})\right)\,s^i\,\sqrt{q}\,\dd \theta\,\dd \varphi\,,
\end{equation}
where a fiducial metric $f_{ij}$ is introduced on $\Sigma_t$ which must be Euclidian, i.e. flat.  Since the integral must be calculated over 2-spheres $\mathscr{S}$, the obvious choice is $f_{ij} = \text{diag}(1,\,r^2,\,r^2 \,\sin^2\theta)$.  In the above expression furthermore $\mathcal{D}_i$ is $f_{ij}$'s Levi-Civita connection, $s^i$ the normal vector induced on the 2-spheres, and $\sqrt{q}$ the volume element with respect to the physical metric $\gamma_{ij}$. Thus, one must take special care to distiguish the quantities that depend on the physical metric, $s^i$ and $\sqrt{q}$, from those that depend on the flat fiducial metric, $f^{ij}$ and $\mathcal{D}_i$. With this in mind, the resulting expression for our particular metric \eqref{eq:spatial_metric} is \cite{Gourgoulhon_2010}
\begin{equation}\label{eq:ADM_mass_developed}
M_{\text{ADM}} = \lim_{r_{\text{bdy}}\to \infty}\,-\frac{1}{8}\,\int\limits_{0}^{\pi}\,\dd\theta\,\left(\pdv{r}\,(A + H) + \frac{H - A}{r}\right)\,r^2\,\sin\theta\,.
\end{equation}

It is well known that the ADM mass converges very slowly with radius. Even for a Schwarzschild black-hole with mass parameter $M$ (in isotropic coordinates so that $A = H = (1 + M/2r)^4$) the expression turns out to be
\begin{equation}\label{eq:ADM_Schwarschild}
M_{\text{ADM}} = \lim_{r\to \infty} M\,\left(1 + \frac{M}{2r}\right)^3 = M + \mathcal{O}(r^{-1})\,.
\end{equation}
In fact, to have a relative error of less than $1\%$ in this expression, we must have the external boundary at $r_\infty > 150 M$, which can be very computationally expensive. Fortunately, alternatives have been derived. In Schwarzschild's case, a pseudo-mass has been introduced by Alcubierre \textit{et al.} \cite{Alcubierre_2000}
\begin{equation}\label{eq:PS_mass}
M_{\text{PS}} = \left(\frac{\tilde{A}}{16\pi}\right)^{1/2}\,\left(1 - \frac{\left(\dv*{\tilde{A}}{r}\right)^2}{16\pi\,\tilde{\gamma}_{rr}\,\tilde{A}}\right)\,,
\end{equation}
where $\tilde{A}$ is the area of a 2-sphere, $r$ is the radial coordinate, and $\tilde{\gamma}_{rr}$ is the average of $\gamma_{rr}$ over the 2-sphere. This expression is ideal for the case of an areal radial coordinate, where $\tilde{A} = 4\pi\,r^2$, so that 
\begin{equation}\label{eq:PS_Schwarschild}
M_{\text{PS}} = \frac{r}{2}\,\left(1 - \frac{1}{\gamma_{rr}}\right)\,.
\end{equation}
For Schwarzschild's case in the areal coordinates we have $\gamma_{rr} = (1 - 2M/r)^{-1}$, and thus we can easily see that the pseudo-Schwarzschild mass gives an exact expression, $M_{\text{PS}} = M$, independent of our radial coordinate $r$. This certainly is an improvement from the $\mathcal{O}(1/r)$ convergence in the ADM case. However, notice that it will not always be possible to work with an areal radial coordinate, especially for spacetimes that are not spherically-symmetric. This is certainly not possible in our case, since $A \neq H$ for rotating boson stars. But even so, we will later show that in the limit in which $A \approx H$ this expression yields a better result than the ADM mass at finite radius.

Better suited to our problem is the Komar mass \cite{Wald_1984}. Here the only added assumptions are that the spacetime has a Killing vector $\xi^\mu$ associated to stationary symmetry, and that the Ricci tensor $R_{\mu \nu} = 8\pi\,(T_{\mu \nu} - g_{\mu\nu} T/2)$ is zero beyond a 2-sphere $\mathscr{S}$. The expression for the Komar mass is
\begin{equation}\label{eq:M_Komar_theoretical}
M_{\text{Komar}} = \frac{1}{8\pi}\,\oint_{\mathscr{S}}\,\nabla^\mu \xi^\nu\,\dd S_{\mu\nu}\,,
\end{equation}
where $\dd S_{\mu\nu}$ is the 2-form normal to the surface $\mathscr{S}$, i.e. induced by $n^\mu$ and $s^\mu$, the normal vector to the hypersurface $\Sigma_t$ and the normal vector to $\mathscr{S}$, so that $\dd S_{\mu\nu} = (s_\mu n_\nu - s_\nu n_\mu)\,\sqrt{q}\,\dd\theta\,\dd\varphi$. 

Our problem does not actually satisfy the fact that $T_{\mu \nu} = 0$ beyond any 2-sphere $\mathscr{S}$. However, this necessity can be relaxed as long as this tensor goes sufficiently rapid to zero at spatial infinity \cite{Wald_1984}. Indeed, the stress energy tensor is quadratic in the scalar field, and we determined previously that the scalar must decay exponentially, so that at sufficiently large $r$, the field is dominated by this decay,
\begin{equation}
T_{\mu \nu} \to e^{-2\sqrt{m^2 - \omega}\,r}\,.
\end{equation}
Therefore, an expression can be calculated in terms of our metric variables as long as we take the limit at spatial infinity, $r_{\text{bdy}} \to \infty$. We find:
\begin{equation}\label{eq:M_Komar_surface}
M_{\text{Komar}} = \lim_{r_{\text{bdy}}\to \infty}\,\frac{1}{2}\int\limits_{0}^{\pi}\,\dd \theta\,\left(\pdv{\alpha}{r} - \frac{H r^2\sin^2\theta}{2\alpha}\,\Omega\,\pdv{\Omega}{r}\right)\,\sqrt{H}\,r^2\sin\theta\,.
\end{equation}

In addition, another expression can be obtained by using Stokes' theorem and Einstein's equations to convert the surface integral into a volume integral. 
\begin{equation}\label{eq:Komar_mass_volume}
M_{\text{Komar}} = 2\,\int_{\Sigma}\,\left(T_{\mu \nu}- \frac{T}{2}\,g_{\mu \nu}\right)\,n^\mu \xi^\nu\,\dd V\,.
\end{equation}
Above, $\Sigma$ is a spatial hypersurface such that its boundary is the 2-sphere $\mathscr{S}$. Like in the previous case, even though $R_{\mu \nu} \neq 0$, if we take the limit at spatial infinity, the expression can hold. In this case, we just integrate over the complete hypersurface $\Sigma_t$. There is, however, another caveat to this expression: it is clear that for a black-hole like Schwarzschild's, $R_{\mu \nu} = T_{\mu \nu} = 0$, yet the mass is certainly not zero. The issue here is the presence of a spacetime singularity which does not allow an application of Stokes' theorem. Thus, \eqref{eq:Komar_mass_volume} must not be used in the presence of singularities. However, all our functions are regular and smooth at all our hypersurfaces by construction, so we can use the Komar mass in terms of a volume integral \cite{Gourgoulhon_2010}, which in our case reduces to:
\begin{equation}\label{eq:M_Komar_volume}
M_{\text{Komar}} = \lim_{r_{\text{bdy}} \to \infty}\,2\pi\,\int\limits_0^{r_{\text{bdy}}}\,\dd r\int\limits_0^\pi\,\dd\theta\,\left(\frac{2\omega}{\alpha}\,(\omega + l \Omega) - \alpha\,m^2\right)\,\phi^2\,A\,\sqrt{H}\,r^2\sin\theta\,.
\end{equation}
This last expression makes explicit the fact that the integral must converge exponentially, since all terms are quadratic in the scalar field. This certainly is a very different behavior than that of the ADM mass. Also, having two expressions for the same quantity can be very useful for numerical convergence-monitoring. 

Nevertheless, \textit{a priori} there is no connection between the ADM mass and the Komar mass, but it has been shown that they are identical if the Killing vector $\xi^\mu$ is orthogonal to $\Sigma_t$ at spatial infinity. This is true in our case since, as we approach spatial infinity we have
\begin{equation}\label{eq:normal_vector}
n^\mu = \frac{1}{\alpha}\,\left((\pdv*{t})^\mu - \beta^\mu\right) \to (\pdv*{t})^\mu = \xi^\mu\,.
\end{equation}

In conclusion, if we obtain a rapidly convergent Komar mass, this is equivalent to calculating the ADM mass. The clear advantage is that the Komar mass may be calculated within a smaller computational domain, which is ideal for our numerical implementation. In other words, as long as the scalar field is appropriately confined within our computational domain, we will be able to calculate spacetime's global mass parameter.

Our spacetime possesses another symmetry, namely axisymmetry under the Killing vector $\chi^\mu = (\pdv*{\varphi})^\mu$. An ADM and Komar analogue quantity may be derived: the angular momentum. Once again, the ADM formalism converges too slowly and a Komar expression is preferred \cite{Wald_1984,Gourgoulhon_2010}.  For our metric the final expression becomes:
\begin{align}
\begin{split}\label{eq:J_Komar_surface}
J_{\text{Komar}} =&\, \lim_{\mathscr{S}\to \infty}\,\frac{1}{16\pi}\,\oint\,\frac{H\,r^2\,\sin^2\theta}{\alpha\,\sqrt{A}}\,\pdv{\Omega}{r}\,\dd A = \lim_{r_{\text{bdy}} \to \infty}\,\frac{1}{8}\,\int\limits_0^\pi\,\dd \theta\,\frac{H^{3/2}\,r^4\,\sin^3\theta}{\alpha}\,\pdv{\Omega}{r}\,,
\end{split}\\
\begin{split}\label{eq:J_Komar_volume}
=&\, l\,\int_{\Sigma_t}\,\frac{(\omega + l\,\Omega)}{\alpha}\,\phi^2\,\dd V = \lim_{r_{\text{bdy}} \to \infty}\,2\pi\,l\,\int\limits_0^{r_{\text{bdy}}}\,\dd r\,\int\limits_0^\pi\,\dd\theta\,\frac{(\omega + l\Omega)}{\alpha}\,\phi^2\,A\,\sqrt{H}\,r^2\sin\theta\,,
\end{split}
\end{align}
where we have written both expressions in terms of surface and volume integrals. Analogously to the previous expression \eqref{eq:M_Komar_volume}, the volume integral makes explicit the fact that the integral will converge exponentially as $r_{\text{bdy}} \to \infty$.

%As a complementary note, and to have a better understanding of the global quantities and their relation the external boundary conditions, we now turn towards the Kerr black-hole spacetime in quasi-isotropic coordinates. This is possible since Kerr is stationary and axisymmetric. We shall only consider the case where there is no electromagnetic charge. It is simple to calculate asymptotic series for the usual $M,\,a = J/M$ parameters
%
%\begin{align}\label{eq:Kerr_behavior}
%\begin{split}
%\aleph =&\, -\frac{M}{r} - \frac{M^2}{r^2} + \mathcal{O}(r^{-3},\,\theta)\,,
%\end{split}\\
%\begin{split}
%\Omega =&\, -\frac{2(aM)}{r^3} + \frac{4M\,(aM)}{r^4} + \mathcal{O}(r^{-5},\,\theta)\,,
%\end{split}\\
%\begin{split}
%a =&\, +\frac{M}{r} + \frac{M^2\,(2 - \sin^2\theta)}{2r^2} + \mathcal{O}(r^{-3},\,\theta)\,,
%\end{split}\\
%\begin{split}
%h =&\, +\frac{M}{r} + \frac{M^2}{r^2} + \mathcal{O}(r^{-3},\,\theta)\,,
%\end{split}\\
%\begin{split}
%\lambda =&\, -\frac{M^2}{r^4} - \frac{4M^3}{r^5} + \mathcal{O}(r^{-6},\,\theta)\,.
%\end{split}
%\end{align}
%
%\noindent This previous set of equations keeps the first two terms for each grid function. Notice, in particular, that the $\aleph$ and $h$ expansion is equal (save from a minus sign) up this order. 

%%%%%%%%%%%%%%%%%%%%%%%%%%%%%%%%%
%%%   GLOBAL NEWTON METHODS   %%%
%%%%%%%%%%%%%%%%%%%%%%%%%%%%%%%%%

\section{Global Newton Methods}\label{sec:Global Newton Methods}

Initially, our system of equations for $\lbrace f_\aleph,\,f_\Omega,\,f_a,\,f_h,\,f_\psi,\,f_\lambda\rbrace$ appears to consist of six elliptic, nonlinear, coupled equations. However, this can be misleading. When we discretize each variable $\lbrace \aleph,\, \Omega,\,a,\,h,\,\psi,\,\lambda \rbrace$ on our grid $(\rho_i,\,z_j)$ we obtain six equations for each grid point. For example, for an illustrative $800$ interior points in each direction, we have $6 \times 800 \times 800 \approx 4\times 10^6$ equations, and each function $\lbrace \aleph(\rho_i,\,z_j),\, \Omega(\rho_i,\,z_j),\, a(\rho_i,\,z_j),\,h(\rho_i,\,z_j),\,\psi(\rho_i,\,z_j), \,\lambda(\rho_i,\,z_j)\rbrace$ is a true variable we must solve for. Adding to the complexity is that we also must calculate the field's frequency, $\omega$, as a sort of eigenvalue. It is easy to see that we do not have enough equations by adding $\omega$ (actually $\xi$ as defined in Equation \eqref{eq:xi} above) as an unknown. The solution is to remove another variable, i.e., constrain  the value of $\psi$ at a certain grid point, which also solves the previous requirement that $\psi > 0$ in order to avoid the trivial Minkowski solution. Of course, we can fix $\psi$ at any point in theory, however, it is best to constrain it near the axis and origin, i.e., the first interior point $\psi(\Delta\rho/2,\,\Delta z/2)$. Otherwise, we might be constraining it at a point where it is already too small given its exponential decay. Due to the axial and equatorial symmetries, this is equivalent to constraining the field's value at the origin:
\begin{equation}
\psi_0\equiv \psi(0,\,0)\,.
\end{equation}
Now, suppose that for each $\psi_0$ (and each rotation number $l$) there exists a unique $\omega$ for the ground-state (the converse is not true, i.e., for a single $\omega$ there can be multiple $\psi_0$'s, as will be seen later in Section \ref{sec:Results}), then the system can be solved with $\omega$ as an unknown and the above constraint $\psi(0,\,0) = \psi_0$.

This is can be done by a Newton-Raphson method \cite{Kollerstrom_1992,Ypma_1984}. The following discussion on Newton methods and their globalization closely follows Deuflhard \cite{Deuflhard_2011}; other globalization strategies have also been developed, for example, by Bank and Rose \cite{Bank_1981}. Specifically, we will use a finite difference discretization \cite{Leveque_2007} and calculate the Jacobian matrix with respect to the discrete variables and $\omega$ (since it is also an unknown variable). Newton's method is a very powerful tool used to solve nonlinear equations, specifically when we have a system of equations from a domain $D \subset \mathbb{R}^n$ into a codomain space $U\subset \mathbb{R}^n$, i.e., for $x \in \mathbb{R}^n$, we seek the solution of the $n$ equations represented as a vector in $\mathbb{R}^n$
\begin{equation}
F(x) = 0\,.
\end{equation}
The standard Newton method relies on an initial guess $x^0$ and a basic linearization of the equation:
\begin{equation}
F(x^k + \Delta x^k) = F(x^k) + F'(x^k)\,\Delta x^k + \mathcal{O}((\Delta x^k)^2)\,,
\end{equation}
where $F'(x^k)$ is the $n\times n$ Jacobian matrix at $x^k$. Thus, Newton affirms that a ``better'' estimate is constructed iteratively as
\begin{equation}
    x^{k + 1} \equiv x^k + \Delta x^k\,,\quad \Delta x^k = -F'(x^k)^{-1}\,F(x^k)\,.
\end{equation}

The (famous) advantage of this method is that it converges quadratically \cite{Kantorovich_1982}, however it can be very sensitive to the initial guess $x^0$. The ``more nonlinear'' the problem, this sensitivity is exacerbated. Unfortunately, for a general problem, we might really have no idea as to what is a ``good initial guess''. Therefore, we would like to globalize Newton's method, so that in the best case scenario any type of initial guess leads to the true solution. This is immediately limited by the fact that a system of equations might have multiple (or none at all) solutions. However, if we constrain our solution appropriately via boundary conditions, we may recover uniqueness. Thus, although a formal proof can be very cumbersome, uniqueness is critical and we will normally ``build-in'' constraints into the very same variables. For example, in our problem we have guaranteed $m^2 > \omega^2$ by introducing the variable $\xi$ \eqref{eq:xi}, and we are also constraining the metric variables to be positive by taking their logarithms. Also, as previously discussed, in looking for the scalar field's ground state, we should enforce $\psi> 0$.

Our first step in this discussion is having the means to determine if a problem requires a globalization, or if the classical Newton algorithm will suffice. In other words, we must determine if a problem is ``highly nonlinear''. First, we notice that the classical existence, uniqueness and convergence theorems \cite{Kantorovich_1982,Ortega_2000} for Newton's method rely on having \textit{a priori}\/ that $F'(x)^{-1}$ exists and is bounded across the domain $D$, 
\begin{equation}\label{eq:d1.2}
\norm{F'(x)^{-1}} \leq \beta < \infty\,, \quad x \in D\,.
\end{equation}
However, this is computationally problematic. First of all, notice that this norm is taken over matrices.  Also, $\beta$ is realistically near impossible to calculate for nontrivial problems. The best we can do is to sample across the domain
\begin{equation}\label{eq:d1.3}
\norm{F'(x^0)^{-1}} \leq \beta_0\,, \quad x^0 \in D\,.
\end{equation}
But even this is still expensive as we have to explicitly calculate the inverse $F'(x^0)^{-1}$ which can be numerically very expensive. The solution is to use another condition instead of \eqref{eq:d1.2}, namely a Lipschitz condition
\begin{equation}\label{eq:d1.4}
\norm{F'(x) - F' (y)} \leq \gamma \, \norm{x - y}\,, \quad x,\,y \in D\,.
\end{equation}
This leads to the Newton-Kantorovich theorem \cite{Kantorovich_1982} that asserts existence and uniqueness of a solution $x^*$, and quadratic convergence for initial guess $x^0$ that is in a neighborhood characterized by the Kantorovich quantity
\begin{equation}
h_0 \equiv \norm{x^* - x^0} \,\beta_0\,\gamma < \frac{1}{2}\,,
\end{equation}
where $\beta_0$ and $\gamma$ were defined in \eqref{eq:d1.3} and \eqref{eq:d1.4}, respectively. However, this does not seem to solve the problem, since we must still somehow calculate the Lipschitz quantity, $\gamma$, and this again implies norms over matrices. Fortunately, a reformulation of the Newton-Kantorovich theorem \cite{Deuflhard_2011} exists which only requires vector norms and another Lipschitz constant, $\epsilon$:
\begin{equation}\label{eq:d1.7}
\norm{F'(x)^{-1}\,\left(F'(x)-F'(y)\right)\,\left(x - y\right)} \leq \epsilon\,\norm{x - y}^2\,, \quad x,\,y \in D\,.
\end{equation}
Furthermore, with this condition, the Kantorovich quantity is written simply as 
\begin{equation}
h_0 \equiv \norm{x^* - x^0}\,\epsilon < 2\,.
\end{equation}
In other words, we have unique quadratic convergence for a neighborhood of radius \cite{Deuflhard_2011}
\begin{equation}\label{eq:convergence_radius}
\rho_0 \equiv \norm{x^* - x^0} = \frac{2}{\epsilon}\,.
\end{equation}

As previously stated, we now have only vector norms in both sides of \eqref{eq:d1.7}, which is computationally advantageous. Still, calculating the Lipschitz constant $\epsilon$ seems to be a hopeless task based on random sampling. However, condition \eqref{eq:d1.7} has an enormous structural advantage in that it is \textit{affine covariant}, i.e., it is invariant under the transformation of the original system of $n$ equations, $F(x) = 0$ into
\begin{equation}
G(x) = A\,F(x) = 0\,,
\end{equation}
where $A \in \text{GL}(n)$ is a nonsingular $n\times n$ matrix. This unique property allows a clever construction of a scalar function $g(x,\,y,\,z)$ such that
\begin{equation}
\epsilon = \text{sup}\,g(x,\,y,\,z)\,, \quad x,\,y,\,z \in D\,.
\end{equation}
Once this function, $g$, is obtained, we can calculate a \textit{computational local estimate}
\begin{equation}\label{eq:comp_available}
\left[ \epsilon_0\right] = g(x^1,\,x^0,\,x^0) \leq \epsilon\,.
\end{equation}
A straightforward example (although not the actual estimate used in algorithms) is \cite{Deuflhard_2011}
\begin{equation}
\left[\epsilon_0 \right] = \frac{\norm{F'(x^0)^{-1}\,\left(F'(x^1)-F'(x^0)\right)}}{\norm{x^1 - x^0}}\,.
\end{equation}

There will obviously be an error $\epsilon - \left[\epsilon_0\right]$, however an efficient adaptive Newton algorithm can be constructed in such a way that successive iterations catch more binary digits of the true $\epsilon$, via the bit counting lemmas as presented by Deuflhard \cite{Deuflhard_1975,Deuflhard_2011}. In summary, Newton's method, the vector norm Lipschitz condition \eqref{eq:d1.7}, and affine covariance gives us a way to calculate an approximate $\left[\epsilon\right]$. Via the convergence radius \eqref{eq:convergence_radius}, we can now estimate whether we have an initial guess ``sufficiently close'' to the solution, $x^*$, i.e., we characterize a problem as \textit{highly nonlinear} \cite{Deuflhard_2011} if the Newton update is such that
\begin{equation}
\norm{\Delta x^0} = x^1 - x^0 \gg \frac{2}{\left[\epsilon_0\right]}\,.
\end{equation}

In such a case, Newton's method will not have guaranteed convergence and requires a globalization. This can only be constructed using additional structure on the equations $F(x)$. A globalization of Newton's method should deal with (within reason) ``bad'' initial guesses and should merge with the classical Newton method when sufficiently near to the solution $x^*$ as determined via Equation \eqref{eq:convergence_radius} since we wish to make use of Newton method's quadratic convergence. 

There exists multiple globalization strategies, however the most intuitive is probably that of steepest descent \cite{Cauchy_1847} where the iterates $x^k$ successively approach the solution point whose direction is calculated via the gradient of the residual level function
\begin{equation}\label{eq:first_level_function}
T(x) \equiv \frac{1}{2}\,\norm{F(x)}^2 = \frac{1}{2}\,F(x)^T\,F(x)\,.
\end{equation}
This is a concave function, since $T(x) = 0$ if and only if $x = x^*$ and $T(x) > 0$ if and only if $x \neq x^*$. Thus, a global solution can easily be found via a \textit{monotonicity criterion}
\begin{equation}\label{eq:mc}
T(x^{k+1}) < T(x^k) \quad \iff \quad \norm{x^{k+1}-x^*} < \norm{x^k - x^*}\,.
\end{equation}
A formal iterative method is thus
\begin{align}
\Delta x^k \equiv&\, \grad{T(x^k)} = -F'(x^k)^T\,F(x^k)\,,\\
x^{k+1} \equiv&\, x^k + s_k\,\Delta x^k\,,\\
s^k >&\, 0\,,
\end{align}
where $s^k$ is a steplength parameter, which is chosen to satisfy the monotonicity criterion \eqref{eq:mc} by taking the successive steps $i = 0,\,1,\,\ldots$ until it is true that
\begin{equation}
T(x^k + s^i_k\,\Delta x^k) < T(x^k)\,,
\end{equation}
via a steplength strategy
\begin{equation}
     s^{i+1}_k \equiv \kappa\, s^i_k\,,\quad \kappa < 1\,,\quad s^0_k = 1\,,
\end{equation}
where $\kappa$ is usually chosen to be $\tfrac{1}{2}$. This method can be shown to have linear convergence \cite{Deuflhard_2011} for arbitrarily ``bad'' initial guesses $x^0$ and it is also clear that as the steplength goes to one, we recover Newton's classical method. However, this convergence, although linear, can be also arbitrarily slow. Worse still is the fact that we can find ``pseudo-convergence'' characterized by small 
\begin{equation}
\norm{F'(x)^T\,F(x)} \ll 1\,.
\end{equation}

This discussion on the method of steepest descent is not fruitless since it serves to introduce the analogous and more powerful extensions where we do not utilize the level function \eqref{eq:first_level_function} but a \textit{general level function} to fully utilize affine covariance \cite{Deuflhard_2011}
\begin{equation}
    T(x\,|A) \equiv \frac{1}{2}\,\norm{A\,F(x)}^2\,,\quad A \in \text{GL}(n)\,.
\end{equation}
Using this general level function, it is easily shown that its minimization still points ``downhill'' independently of the matrix $A$ (notice that $A = I$ gives the steepest descent method). 

Several choices of matrices are possible to implement different methods, including the trust-region models of Levenberg-Marquardt \cite{Levenberg_1944,Marquardt_1963}. The choice of $A = F'(x^k)^{-1}$ gives the \textit{error oriented descent} where the iterates now are required to satisfy the \textit{natural monotonicity test} \cite{Deuflhard_1975,Deuflhard_2011} 
\begin{equation}\label{eq:natural_mono}
    \norm{\overline{\Delta x}^{k+1}} < \norm{\Delta x^k}\,,
\end{equation}
where we introduce the \textit{simplified} Newton correction
\begin{equation}
    \overline{\Delta x}^{k+1} \equiv F'(x^k)^{-1}\,F(x^{k+1})\,.
\end{equation}
Furthermore, we now interpret $s_k$ as a \textit{damping factor}
\begin{equation}
    x^{k+1} = x^k + s_k\,\Delta x^{k+1}\,,
\end{equation}
that must be chosen via a damping strategy that obeys the criteria \eqref{eq:natural_mono} and goes to one once we are within the local convergence radius \eqref{eq:convergence_radius}. This damping strategy is given by Theorem 3.12 of Deuflhard \cite{Deuflhard_2011} and states that given a Lipschitz condition \eqref{eq:d1.7} and a Kantorovich quantity
\begin{equation}
    h_k \equiv \norm{\Delta x^k}\,\epsilon\,,
\end{equation}
the natural monotonicity criteria \eqref{eq:natural_mono} is satisfied and the optimal choice is
\begin{equation}
s_k \equiv \min(1,\,1/h_k)\,.    
\end{equation}

Deulfhard also gives a proof for global convergence in Theorem 3.14 of his book \cite{Deuflhard_2011}. Our only remaining limitation is that this theoretical damping strategy cannot be implemented directly and only via the computational available estimates $[\epsilon_k]$ introduced in Equation \eqref{eq:comp_available} and $[h_k] = [\epsilon_k]\,\norm{\Delta x^k}$. The supremum property implies that 
\begin{equation}
[s_k] \equiv \min(1,\,1/[h_k]) \geq s_k\,,
\end{equation}
so that the estimated damping factor may be ``too large''. This conundrum is addressed with a \textit{prediction and correction strategies}. In essence, these strategies use trial iterates and a user-input of an initial damping factor $s_0$ to guarantee the monotonicity criteria. These are also termination criteria in the algorithm since the damping factor may become arbitrarily small for singular problems. More details on these criteria and their optimization is presented in Deuflhard \cite{Deuflhard_2011} where the complete implementation of these concepts results in the error oriented global Newton method, called NLEQ-ERR \cite{NLEQ1,Deuflhard_2011}, that serves as our solution algorithm presented in Appendix \ref{sec:Appendix 2}.

%%%%%%%%%%%%%%%%%%%%%%%%%
%%%   CODE OVERVIEW   %%%
%%%%%%%%%%%%%%%%%%%%%%%%%

\section{Code Overview}\label{sec:Code Overview}

All previous functionalities are implemented in a numerical code specially tailored to this problem, \textit{ROTBOSON}, openly available at a GitHub repository \cite{ROTBOSON}. \textit{ROTBOSON} is a finite difference discretization code spanning over 20,000 lines of C++ OpenMP \cite{OpenMP_1998} code influenced by other cousins \textit{OllinAxis} \cite{Torres_2016} and \textit{OllinBrill} \cite{Ontanon_2018} developed at the ICN, UNAM. Its design philosophy centers on being able to run on personal or professional-grade machines and as such it is scalable and optimized. Its main usage is detailed in the GitHub repository but is briefly summarized in the interest of the following discussion.

First, the user must compile the code with either the GNU or Intel C++ compiler. An installation of the freely available Intel Math Kernel Library software \cite{MKL_PARDISO_2018} is also necessary since it used for the sparse direct solver and other vector operations. Once compiled, the user must provide a parameter file with a list of relevant parameters such as the grid resolutions $\Delta \rho,\, \Delta z$; the number of interior points $N_\rho,\, N_z$; the finite differentiation order (2 or 4); the scalar field's rotation number $l$, an initial guess for the scalar frequency $\omega_0$; and other I/O options to read or interpolate previously obtained datasets. A selection of parameters inside a typical parameter file is presented below.

\begin{lstlisting}[language=Python]
# GRID
dr         = 0.015625 # Step size in rho (r) direction.
dz         = 0.015625 # Step size in z direction.
NrInterior = 512      # Number of interior points in r direction.
NzInterior = 512      # Number of interior points in z direction.
order      = 4        # Finite difference discretization order 2|4.

# SCALAR FIELD PROPERTIES
l = 4   # Rotation number.
m = 1.0 # Field mass parameter.

# INITIAL DATA I/O.
readInitialData = 2 # Select no initial data 0, direct read 1, interpolation 2.
# File names for variables.
log_alpha_i = "./log_alpha_i.asc"
beta_i      = "./beta_i.asc"
log_h_i     = "./log_h_i.asc"
log_a_i     = "./log_a_i.asc"
psi_i       = "./psi_i.asc"
lambda_i    = "./lambda_i.asc"
w_i         = "./w_i.asc"

# FIXED VARIABLE.
fixedPhi  = 1 # Select if scalar field is fixed and if so, where.
fixedPhiR = 2
fixedPhiZ = 2

# SOLVER PARAMETERS
solverType    = 1       # Select Newton solver: 0 classical, 1 NLEQ-ERR, 2 NLEQ-RES.
localSolver   = 1       # Enable local solver.
epsilon       = 1.0E-10 # Convergence exit tolerance.
maxNewtonIter = 20      # Maximum number of iterations.
lambda0       = 1.0E-03 # Initial damping factor.
lambdaMin     = 1.0E-06 # Minimum damping factor.
useLowRank    = 1       # Use Low-Rank optimization 1, otherwise use CGS preconditioning.

# Parameter file continues...
\end{lstlisting}

A critical element of this code is its sparse direct solver (the theory behind these solvers is beyond the scope of this work but may be consulted in Tewarson \cite{Tewarson_1973}, and Gupta and Kumar \cite{Gupta_1995}, among others). Indeed, about 50\% of its lines consist of generating Jacobian matrices at different discretization orders (this was done automatically via Python and Mathematica \cite{Mathematica_2018} scripts). Once the matrices are written in sparse CSR format they are used in Intel MKL's included software, PARDISO \cite{Schenk_2018,PARDISO_1,PARDISO_2}. Since Newton's method solves iteratively, the solver can be optimized greatly given the fact that succesive Jacobians have the same sparse structure, thus an $LU$ sparse factorization scheme only requires a single analysis phase which can then be saved for other executions with entirely different parameters (such as rotation number and step size) but identical sparse structure. For more details on PARDISO's optimization for problems in Numerical Relativity, see Ontanon \cite{Ontanon_2018}.

In terms of memory use and scalability, \textit{ROTBOSON} has been tested succesfully at a cluster node running on up to 54 processing units and with a grid size of approximately $1000\times 1000$ points, \textit{ROTBOSON} uses about 120 Gb of memory. For a personal machine, adequate results can be obtained over 4 processing units and a grid of $400 \times 400$ which uses instead 12 Gb of RAM memory. For now, execution is limited to a single node since the sparse solver loses significant speed when forced to run under a message passing interface, such as MPI \cite{MPI_1994}, over multiple nodes.

%%%%%%%%%%%%%%%%%%%%%%%%%%%%
%%%   ERROR INDICATORS   %%%
%%%%%%%%%%%%%%%%%%%%%%%%%%%%

\section{Error Indicators}\label{sec:Error Indicators}

Before reporting any results, we must verify that our numerical implementation converges appropriately. Since we are using finite differences, we should expect that they introduce truncation errors. For all the following results, finite differences were implemented to fourth order (although second order is also supported for faster performance and testing). Furthermore, the exit tolerance for Newton's algorithm is always set to a relative error $\epsilon < 10^{-10}$, i.e., in terms of the language of Section \ref{sec:Global Newton Methods}, by taking 2-norms, one of these conditions must be true to exit and guarantee a solution $x^*$
$$\frac{\norm{\Delta x^k}}{\norm{x^k}} < 10^{-10}\,,\quad \text{or}\quad\, \frac{\norm{\overline{\Delta x^k}}}{\norm{x^k}} < 10^{-10}\,,$$

To check for fourth order convergence we vary the resolutions $\Delta \rho = \Delta z$, but we must fix the external boundary $r_{\text{bdy}}$ in all solutions. For example, we consider eight different configurations with interior points
$$N_\rho = N_z \in \lbrace 128,\, 256,\, 384,\, 512,\, 640,\,768,\, 896,\, 1024\rbrace\,,$$
and $\Delta \rho = \Delta z$ chosen such that in all cases we have $r_{\text{bdy}} = 128\,(\hbar/m)$ according to Equation \eqref{eq:discrete_coordinates}. Furthermore, since we are fixing the position of the boundary, this is equivalent to constricting $\omega$, and in our tests we take $\omega = 0.7\,(m/\hbar)$. Now, given these eight solutions, we must compare quantities to test for fourth order convergence. A possibility is to interpolate the coarse resolutions into the finest and make a point-wise comparison. However, to avoid compounding interpolation errors and to simplify, we can make convergence tests with global quantities, namely $M_{\text{Komar}}$ and $J_{\text{Komar}}$ in their surface and volume integral expressions (Equations \eqref{eq:M_Komar_surface}, \eqref{eq:M_Komar_volume}, \eqref{eq:J_Komar_surface}, \eqref{eq:J_Komar_volume}). Another error indicator is the scalar field's value at the origin $\psi_0$, obtained by interpolation at each solution. 

Using these indicators and varying resolutions, we compare, i.e. subtract in absolute value, two quantities at neighboring resolutions. For example, for two resolutions $\Delta \rho_0 > \Delta \rho_1$ and their reported masses $M(\Delta \rho_0), \,M(\Delta \rho_1)$ we define the relative error as
\begin{equation}
\epsilon(\Delta \rho_1) \equiv \epsilon(M,\Delta \rho_0, \Delta \rho_1) = \abs{M(\Delta\rho_1) - M(\Delta \rho_0)}\,,
\end{equation}

\begin{figure}
\centering
\begin{minipage}{.48\textwidth}
	\captionsetup{width = 0.9\linewidth}
	\centering
	\includegraphics[width = \linewidth]{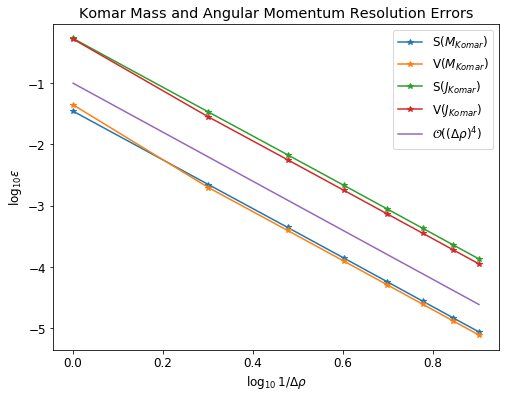}
	\captionof{figure}{Komar mass and angular momentum differences between resolutions as error indicators for $l = 6$, $\omega = 0.7\,(m/\hbar)$, at fixed $r_{\text{bdy}} = 128\,(\hbar/m)$ and varying resolution $\Delta \rho$.}
	\label{fig:error_res1}
\end{minipage}%
\begin{minipage}{.48\textwidth}
	\captionsetup{width = 0.9\linewidth}
	\centering
	\includegraphics[width = \linewidth]{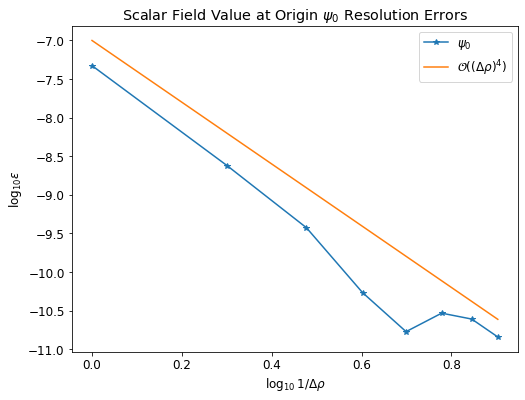}
	\captionof{figure}{Scalar field value at origin $\psi_0$ difference between resolutions as error indicator for $l = 6$, $\omega = 0.7\,(m/\hbar)$, at fixed $r_{\text{bdy}} = 128\,(\hbar/m)$ and varying resolution $\Delta \rho$.}
	\label{fig:error_res2}
\end{minipage}
\end{figure}

Our convergence results are shown as $\log_{10}$-$\log_{10}$ plots in Figures \ref{fig:error_res1} and \ref{fig:error_res2}. In Figure \ref{fig:error_res1} an ``S'' or ``V'' denotes whether the expression was calculated as a surface or a volume integral. Both Figures include a reference straight line for $\mathcal{O}((\Delta\rho)^4)$ fourth order convergence. From the Figures it is clear that we have the expected fourth order convergence. In the case of $\psi_0$, the error stagnates at about $\varepsilon \sim 10^{-11}$, which serves to indicate that point-wise it has reached a limit due to machine round-off error instead. 

Now that we have determined fourth-order convergence, we can find an error approximation via Richardson extrapolation \cite{Richardson_1911} or our global quantities. Thus, at the finest resolution with $\Delta \rho = 0.125\,(\hbar/m)$ and $N_\rho = 1024$, we find for this example
$$M_{\text{Komar}} = (5.550074 \pm 0.000002)\,(m_P^2/m)\,\quad \text{and} \quad J_{\text{Komar}} = (37.1475 \pm 0.0001)\,(m_P^2/m)^2\,,$$
which correspond to relative errors of $4\times 10^{-5}\,\%$ and $3\times 10^{-4}\,\%$, respectively.

\begin{figure}
\centering
\captionsetup{width = 0.65\linewidth}
\includegraphics[width = 0.65\textwidth]{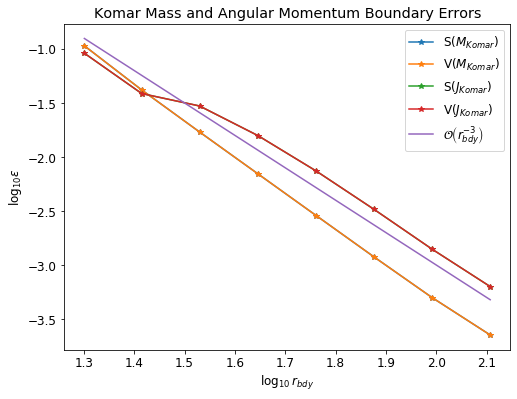}
\caption{Komar mass and angular momentum as error indicators for $l = 6$, $\omega = 0.7\,(m/\hbar)$, for fixed resolution $\Delta \rho = 0.125\,(\hbar/m)$ and varying boundary position $r_{\text{bdy}}$.}\label{fig:error_ndy1}
\end{figure}

Further analysis and testing is prudent, specifically with respect to varying the position of the boundary. In this case, we expect that error arises from the Robin-type boundary conditions such as Equation \eqref{eq:robin_bc} where the right-hand side was not zero, but rather $\mathcal{O}(r_{\text{bdy}}^{-(n+1)})$ for an $n$-type decay. Since the lapse and metrics have $n = 1$, an initial estimate might lead us to expect that there is second order convergence with respect to $r_{\text{bdy}}$. To verify this, other convergence tests are done now with fixed resolution $\Delta \rho = \Delta z = 0.125\,(\hbar/m)$ and varying $r_\text{bdy}$ (which is equivalent to varying the number of interior points). Once again, global quantities are used as error monitors and are presented in Figure \ref{fig:error_ndy1} as $\log_{10}$-$\log_{10}$ plots.

In this case, we can conclude that the errors due to the boundary still converge. By adding the $\mathcal{O}(r_{\text{bdy}}^{-3})$, the results indicate that we have third order convergence instead of the initially expected second order. However, given the complexities of this analysis, it would be best to not read too much into these results and instead just conclude that the error is indeed decreasing as we place the boundary further away, even though the magnitude of the boundary error is at least an order of magnitude greater than the resolution error (in this example, the boundary error is about 30 times greater than the resolution error). It is clear that further analysis and studies are necessary for boundary problems and approximations in Numerical Relativity, but this is outside our current scope.

%%%%%%%%%%%%%%%%%%%
%%%   RESULTS   %%%
%%%%%%%%%%%%%%%%%%%

\section{Results for \texorpdfstring{$l\in [0,\,6]$}{l = [1, ... , 6]}}\label{sec:Results}

\begin{figure}
\centering
\captionsetup{width = 0.95\linewidth}
\includegraphics[width = 0.95\textwidth]{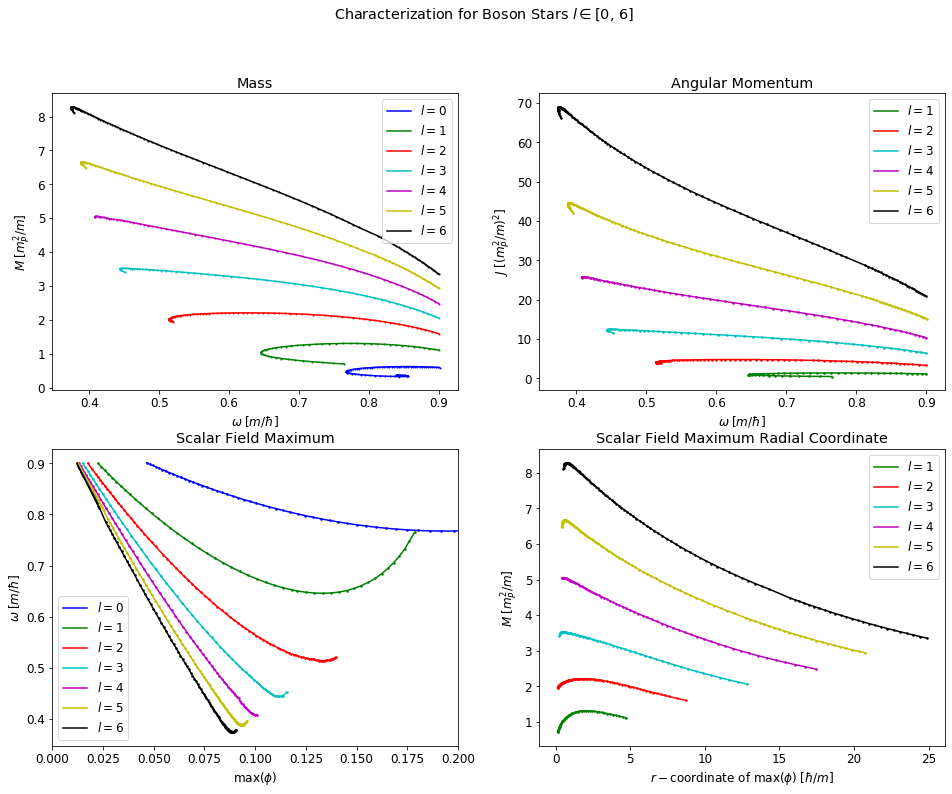}
\caption{Rotating boson stars' characterization for $l \in [0,\,6]$. Included global parameters are the Komar mass, the Komar angular momentum, the field's frequency $\omega$, the field's maximum $\max(\phi)$, and the radial coordinate of this maximum.}\label{fig:results1}
\end{figure}

Having verified that the numerical implementation converges correctly, we now present our results for $l \in [0,\,6]$, which are summarized in Figure \ref{fig:results1}. All solutions therein have a maximum $\omega = 0.9\,(m/\hbar)$. Most solutions are done with grids of $800\times 800$ interior points and $\Delta \rho = \Delta z = 0.04\,(\hbar/m)$. However, for $l = 1,2$ a finer resolution of $\Delta \rho = \Delta z = 0.02\,(\hbar/m)$ is necessary to examine the high-amplitude behavior. Conversely, for $l =5,6$ a coarser resolution of $\Delta \rho = \Delta z = 0.08\,(\hbar/m)$ is required at low amplitudes $\omega > 0.8\, (m/\hbar)$ for which the field's width increases, and in order to properly fit it in our computational domain (given our computational resources) the boundary is placed farther away and the resolution is decreased. The changes in resolution have been done seamlessly, i.e. where the difference between the global parameters is less than the truncation error.

Notice that the spherically symmetric case is also included for reference as $l=0$. In this case the solution is found using an analogous method, but it is worth pointing out that they are not simply obtained by setting $l=0$ in our elliptic equations. For more information on our results for spherical boson stars see \cite{SPHBOSON}. Spherical boson stars \cite{Liebling_2017} trivially have zero angular momentum, $J = 0$, and the field's maximum is always located at the origin \cite{Grandclement_2014}.

Some important maximums and minimums are presented in Table \ref{table:results_table}. Given our error analysis, and considering the worst-case scenario where the error's main contributors are the boundary conditions, we can still report five significant figures for masses and frequencies, and four significant figures for angular momenta. These critical points were obtained by interpolating a fourth-degree spline into the data points and calculating the derivative's zeros. These results are in agreement for $l = 0,1,2$ with those given in references \cite{Grandclement_2014,Liebling_2017,Lai_2004,Schunck_2003,Mielke_2016}. From Table \ref{table:results_table} we also can report new ``turning-points'' (i.e. the minimum value of $\omega$) and maximum masses which, to our knowledge, are not yet present in the literature.

\begin{table}[H]
\centering
\begin{tabular}{c|c|c|c}
Boson Star $l$ & Maximum $M$ ($m_{\text{P}}^2/m$) & Maximum $J$ $(m_{\text{P}}^2/m)^2$ & Minimum $\omega$ $(m/\hbar)$ \\ 
\hline 
\hline
0 & 0.63300 & 0 & 0.76754\\
1 & 1.3155 & 1.382 & 0.64561 \\ 
2 & 2.2159 & 4.810 & 0.51657 \\ 
3 & 3.5287 & 12.49 & 0.44339 \\ 
4 & 5.0590 & 25.83 & 0.40756 \\ 
5 & 6.6681 & 44.63 & 0.38819 \\ 
6 & 8.2824 & 69.02 & 0.37391 \\ 
\end{tabular} 
\caption{Critical points for boson stars $l \in [0,\,6]$.}\label{table:results_table}
\end{table}

\begin{figure}[H]
\centering
\includegraphics[width = \textwidth]{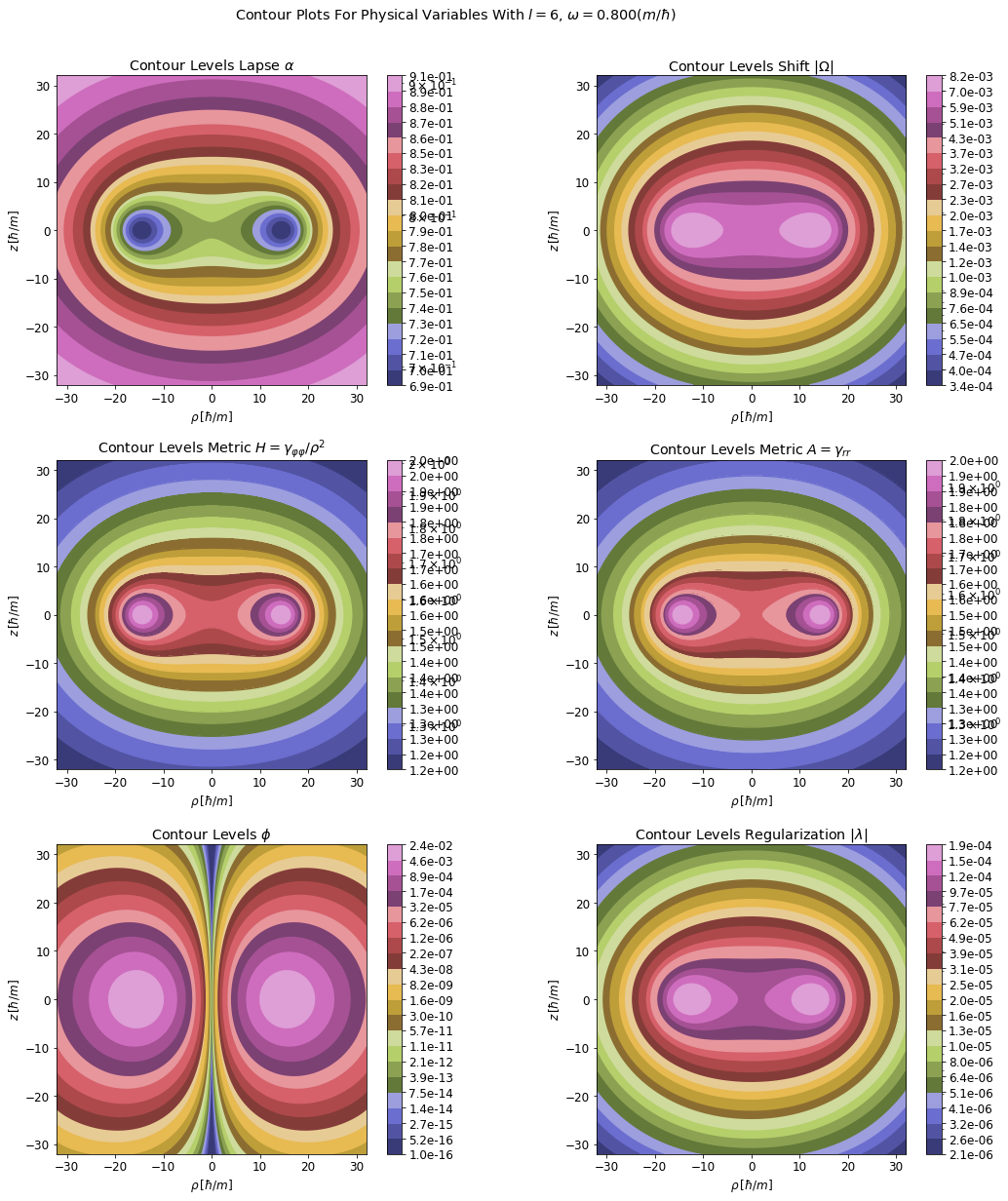}
\caption{Rotating boson star with $l = 6$ for ``low-amplitude'' $\omega \approx 0.8\,(m/\hbar)$.}\label{fig:results2}
\end{figure}

\begin{figure}[H]
\centering
\includegraphics[width = \textwidth]{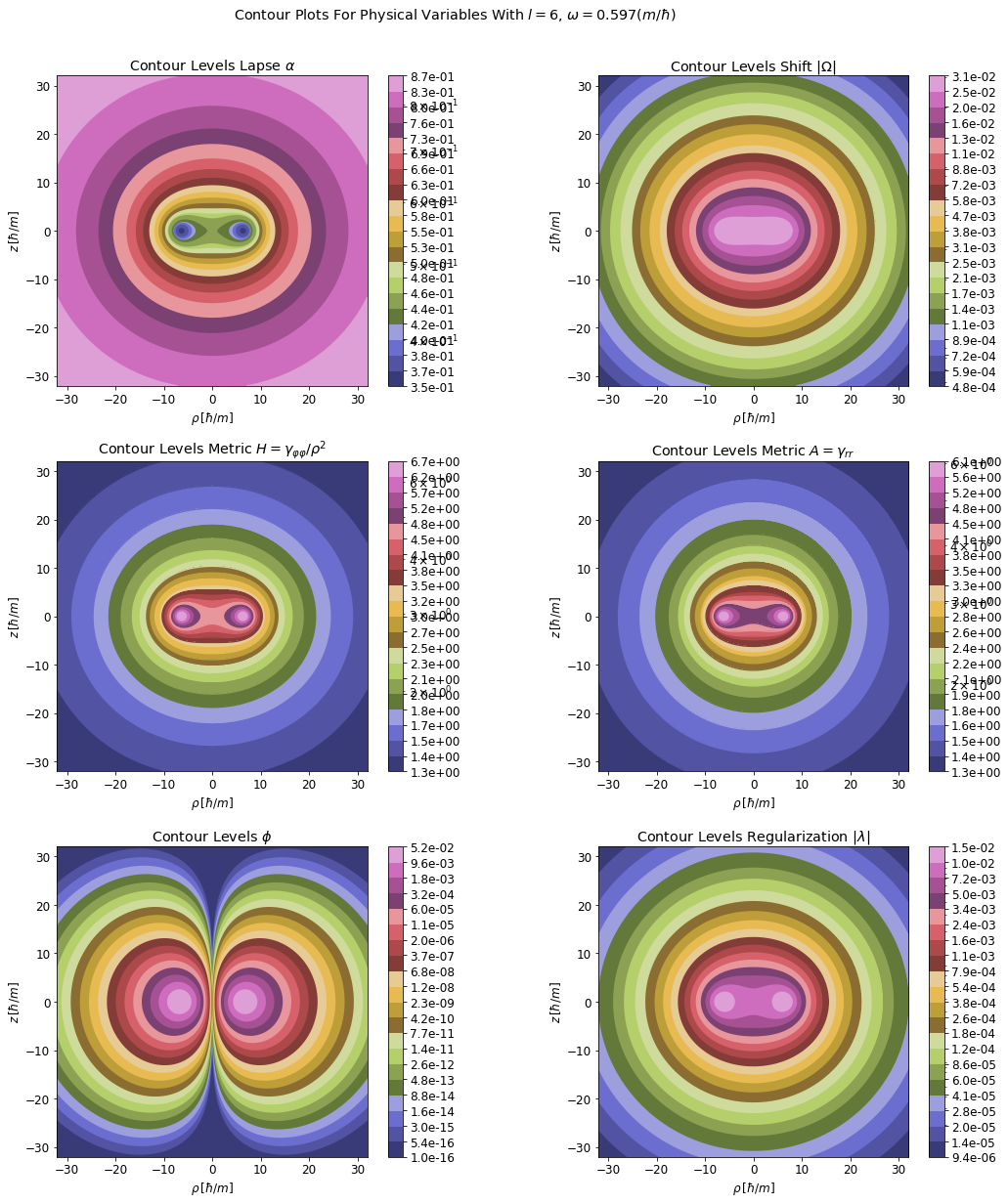}
\caption{Rotating boson star with $l = 6$ for ``medium-amplitude'' $\omega \approx 0.6\,(m/\hbar)$.}\label{fig:results3}
\end{figure}

\begin{figure}[H]
\centering
\includegraphics[width = \textwidth]{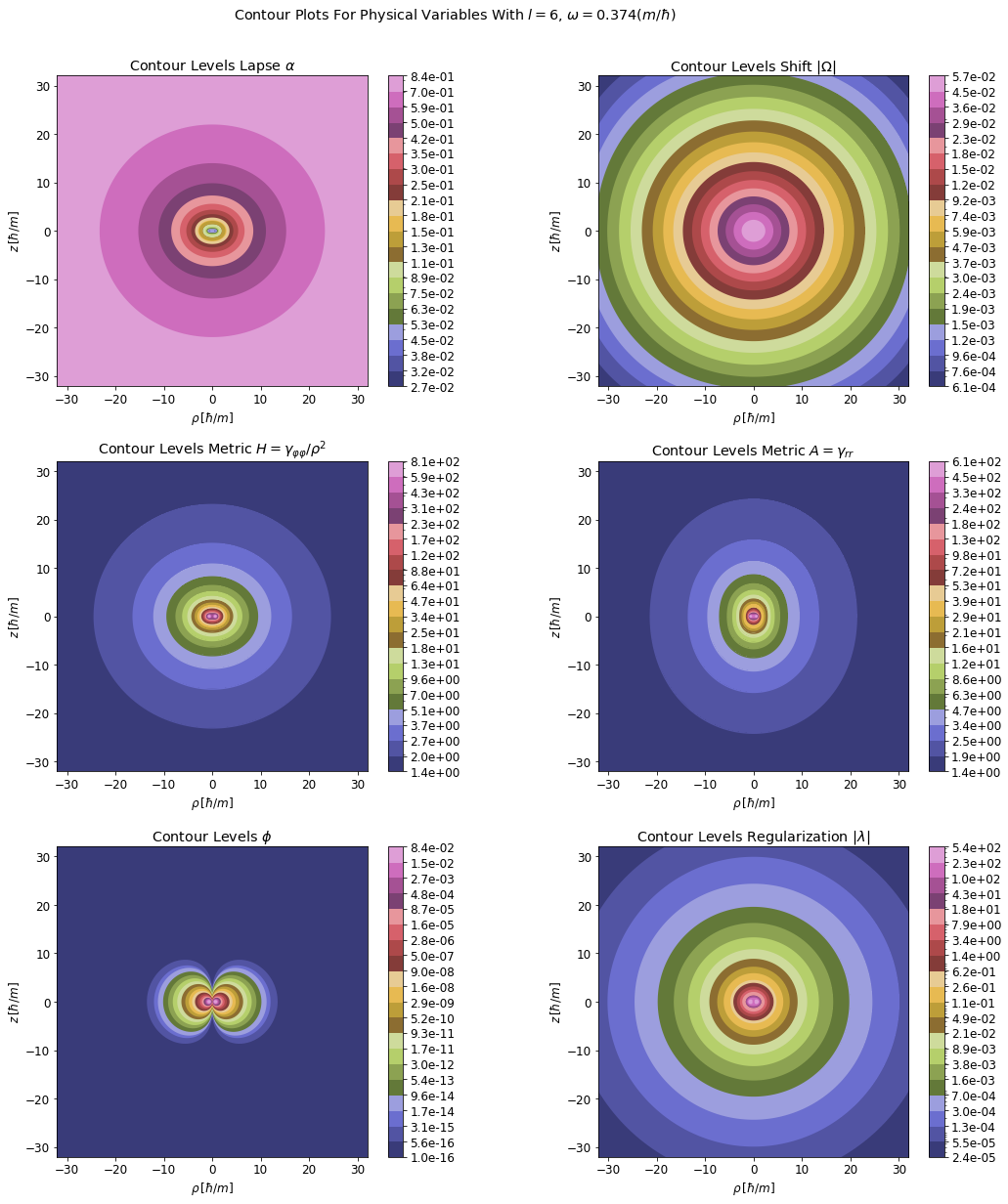}
\caption{Rotating boson star with $l = 6$ for ``high-amplitude'' $\omega \approx 0.4\,(m/\hbar)$.}\label{fig:results4}
\end{figure}

Out of the hundreds of computed data sets, we turn now to examine a couple of illustrative examples for the $l = 6$ case at three different frequencies $\omega \approx \lbrace 0.8\,(m/\hbar),\, 0.6\,(m/\hbar),\,0.37\,(m/\hbar) \rbrace$. These correspond, respectively, to ``low'', ``medium'', and ``high'' scalar field amplitudes. The latter case is at the turning point for $l = 6$ reported in Table \ref{table:results_table}. These results are presented as contour plots for the physical variables $\lbrace \alpha,\,\Omega,\,H,\,A,\,\phi,\,\lambda\rbrace$ in Figures \ref{fig:results2}, \ref{fig:results3}, \ref{fig:results4}. Although, as previously stated, equatorial and axial symmetry imply that our computational domain can be restricted to the quadrant $\rho,\,z > 0$, the figures are plotted in the full space by reflecting about the corresponding axes. A three-dimensional picture can be obtained or imagined by rotating the $z$ axis. Notice that the color bar scale is not equal between figures.

The contours show that, as we decrease $\omega$, the field's spatial extension shrinks and at the same time that its amplitude grows. This behavior is also manifest in the physical variables, namely the metric coefficients $\Omega,\,H,\,A,\,\lambda$, with an opposite behavior for the lapse (i.e. the lapse decreases as the other metric functions grow). Notice that in the ``high-amplitude'' case the central value of the lapse is below $3\times 10^{-2}$.

\begin{figure}[H]
\centering
\captionsetup{width = 0.95\textwidth}
\includegraphics[width = 0.95\textwidth]{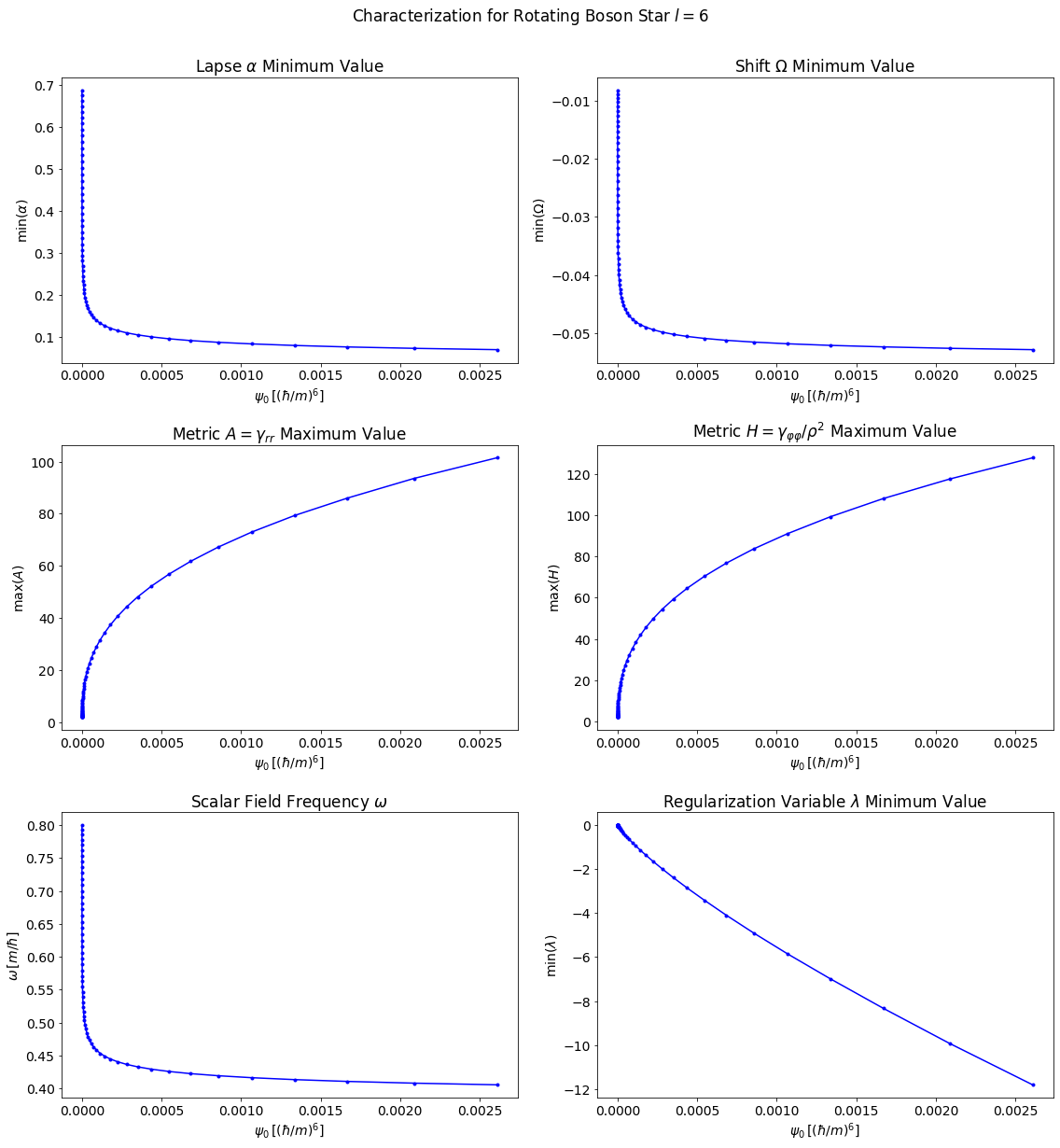}
\caption{Rotating boson stars with $l = 6$. We show the behaviour of the physical variables as functions of $\psi_0$ for the range $\psi_0 \in [1.0\times 10^{-8},\,3\times 10^{-3}]\, (\hbar / m)^6$, at resolutions $\Delta \rho = \Delta z = 0.04\,(\hbar/m)$ and $r_{\text{bdy}} = 32\,(\hbar/m)$.}\label{fig:results7}
\end{figure}

\begin{figure}[H]
\centering
\captionsetup{width = 0.95\textwidth}
\includegraphics[width = 0.95\textwidth]{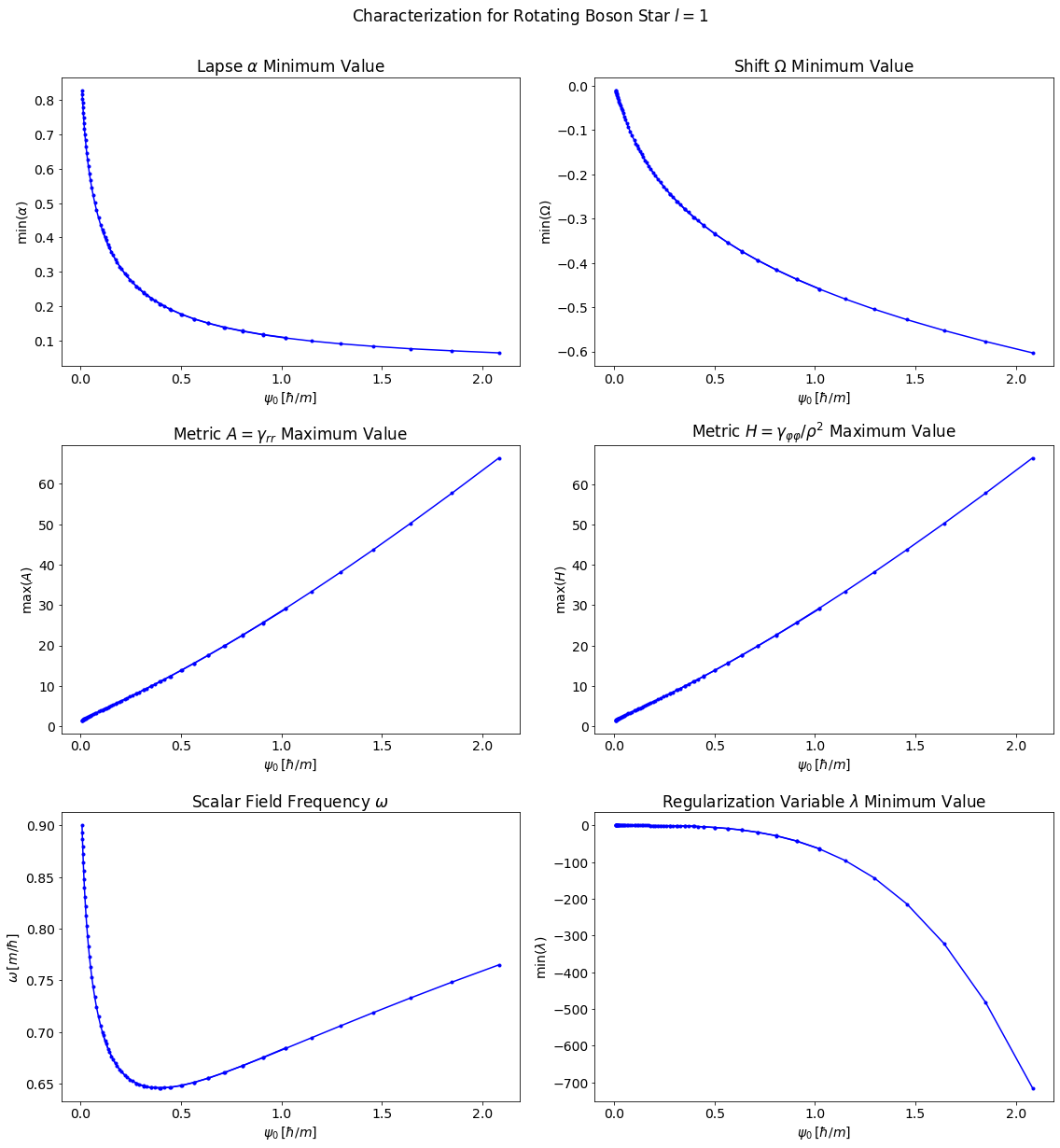}
\caption{Rotating boson stars with $l = 1$. We show the bahavious of the physical variables as functions of $\psi_0$ for the range $\psi_0 \in [1.0\times 10^{-2},\,2.0]\, (\hbar / m)$, at resolutions $\Delta \rho = \Delta z = \lbrace 0.02,\,0.04\rbrace\,(\hbar/m)$ and $r_{\text{bdy}} = 32\,(\hbar/m)$.}\label{fig:results8}
\end{figure}

Considering our three bosons stars with different amplitudes, the results might seem to indicate variable trends as a function of decreasing $\omega$. However, caution is necessary since $\omega$ is not monotonic as a function of $\psi_0$. Rather, as previously discussed the solutions are obtained by varying $\psi_0$. Therefore, an improved picture is obtained in Figure \ref{fig:results7} that shows how the variables behave as we vary this parameter (which has dimensions $(\hbar/m)^l$ given that $\phi$ is dimensionless). This behavior is also studied for the $l = 1$ case in Figure \ref{fig:results8}. These two figures show the true trends of the physical variables. As $\psi_0$ increases, the lapse's minimum decreases, whereas $(-\Omega),\,A,\,H,\,(-\lambda)$ increase. The field's frequency behavior is not monotonic, and its critical points lead to the turning points in the $M$ vs. $\omega$ diagrams.

\begin{figure}[H]
\centering
\captionsetup{width = 0.5\textwidth}
\includegraphics[width = 0.5\textwidth]{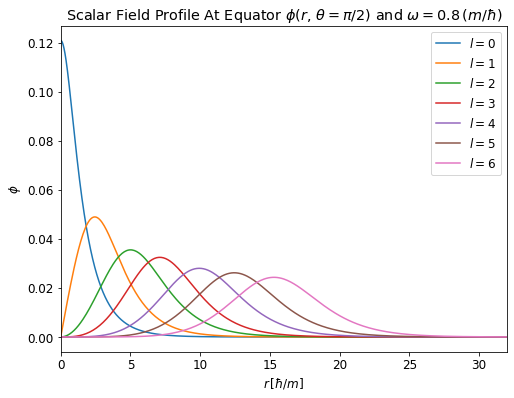}
\caption{Boson stars' scalar field profiles at the equator for fixed $\omega = 0.8\,(m/\hbar)$ and $l \in [0,6]$.}\label{fig:results6}
\end{figure}

The field's localization behavior is also examined for fixed $\omega = 0.8\,(m/\hbar)$ and varying rotation number $l$ in Figure \ref{fig:results6}. Clearly, only at $l = 0$ the maximum may be present at the origin. This figure also shows that an increase in rotation number brings a decrease in the field's amplitude and another increase in terms of the field's extension.

\begin{figure}[H]
\centering
\begin{minipage}{.48\textwidth}
	\captionsetup{width = 0.9\linewidth}
	\centering
	\includegraphics[width = \linewidth]{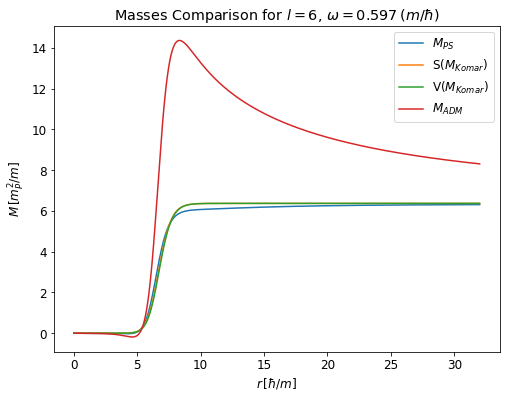}
	\captionof{figure}{Masses comparison and convergence along radial coordinate $r$ for Komar surface, Komar volume, ADM and pseudo-Schwarzschild expressions.}
	\label{fig:results5}
\end{minipage}%
\begin{minipage}{.48\textwidth}
	\captionsetup{width = 0.9\linewidth}
	\centering
	\includegraphics[width = \linewidth]{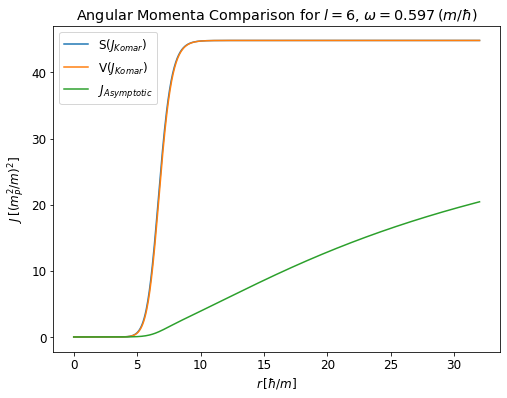}
	\captionof{figure}{Angular momenta comparison and convergence along radial coordinate $r$ for Komar surface, Komar volume and the asymptotic expression.}
	\label{fig:results10}
\end{minipage}
\end{figure}

As previously discussed, we prefer the Komar mass instead of the ADM mass because of the latter's slow convergence. Indeed, Figure \ref{fig:results5} shows how the ADM mass from Equation \eqref{eq:ADM_mass} converges too slowly at \mbox{$r_{\text{bdy}} = 32\,(\hbar/m)$}. This is heavily contrasted with the Komar equivalents which converge exponentially. Notice that the difference between the ADM and Komar masses is about 50\% at $r_{\text{bdy}}$, making the former useless for our results. An analogous results is presented in Figure \ref{fig:results10} for angular momenta where the Komar expressions Equations \eqref{eq:J_Komar_surface} and \eqref{eq:J_Komar_volume} are compared to the asymptotic form 
\begin{equation}
J_{\text{Asymptotic}} = \lim_{r_{\text{bdy}} \to \infty}\,\frac{1}{8}\,\int\limits_0^\pi\,\dd \theta\,r^4\,\sin^3\theta\,\pdv{\Omega}{r}\,.
\end{equation}

\begin{figure}[H]
\centering
\captionsetup{width = 0.65\textwidth}
\includegraphics[width = 0.65\textwidth]{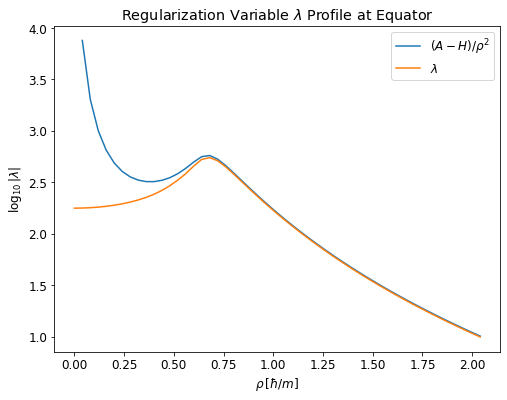}
\caption{Comparison between the regularization variable $\lambda$ and its expression in terms of metric variables $(A-H)/\rho^2$. Both expressions are plotted in absolute value using a logarithmic scale. Notice the considerable differences near the origin, and how they match beyond $r \gg 1\,(\hbar/m)$.}\label{fig:results11}
\end{figure}

The final result of this section is Figure \ref{fig:results11}, which demonstrates the importance of regularization for a ``high-amplitude'' case such as $l = 6$, $\omega = 0.37\,(m/\hbar)$. The yellow curve represents the regularization variable $\lambda$, whereas the blue curve is the expression $(A - H)/\rho^2$ calculated using the two other variables $A,\,H$. Notice that the vertical scale is logarithmic so that at the origin we have $\lambda \approx  2\times 10^2\,(m/\hbar)^2$, whereas $(A - H)/\rho^2$, though indeterminate at the origin, at the first interior point is about $8\times 10^4\,(m/\hbar)^2$ which is an enormous difference of two orders of magnitude. Beyond $r \gtrsim 1\,(\hbar/m)$ both expressions match and can be used interchangeably. However,  $r \ll 1\,(\hbar/m)$ is precisely the region where the scalar field is localized and of non-negligible amplitude. Thus, due to the $-l^2\,\lambda\,\psi/H$ term in Equation \eqref{eq:f_psi}, it is clear that a solution scheme with regularization will have substantial advantages over one without regularization. In our experience, we found that without regularization our solver cannot converge at high amplitudes precisely for this reason. Therefore, thanks to regularization, it has been possible to further explore rotating boson stars solutions at higher amplitudes.

\pagebreak

\pagebreak

\pagebreak

\pagebreak

\pagebreak

%%%%%%%%%%%%%%%%%%%%%%%%%%%%%%%%%%%
%%%   SUMMARY AND CONCLUSIONS   %%%
%%%%%%%%%%%%%%%%%%%%%%%%%%%%%%%%%%%

\section{Summary and Conclusions}\label{sec:Summary and Conclusions}

In this paper we have studied solutions for rotating boon stars, i.e. a stationary and axisymmetric ground-state of the Klein-Gordon complex scalar field coupled to Einstein's equations. The ADM equations initially result in a system of five coupled and highly-nonlinear equations with an added eigenvalue problem for the field's time frequency, $\omega$. Guided by boundary conditions at both spatial infinity and the $\rho,\,z$ axes, we propose variables that mostly automatically satisfy the system's physical constraints. Given the curvilinear formulation of our variables, terms with powers of $1/\rho$ appear throughout the equations.  While most these terms are regular under parity conditions, a single term in the scalar field's equations proportional to $\lambda = (A-H)/r^2$ is not immediately regular. Using local flatness this term is shown analytically regular, but $\lambda$ must be promoted numerically to a new independent variable to guarantee regularity at the rotation axis: thus a sixth elliptic equation is introduced in conjunction with its proper boundary conditions.

For solving our system of equations we have developed an open-source numerical code, \textit{ROTBOSON}, based on finite difference fourth-order discretization and a Cartesian grid in the cylindrical coordinates $(\rho,\,z)$.  This set up is deal for curvilinear axisymmetric spacetimes where the Killing vectors $\xi^a = (\pdv*{t})^a$ and $\chi^a = (\pdv*{\varphi})^a$ are adapted to the coordinates.
Due to the high non-linearity and lack of simple initial guesses to input into a regular Newton-Raphson method, an alternative Global Newton Method using error oriented affine covariant Lipschitz conditions is implemented to solve for an initial seed for each family of rotating boson stars. From this seed, the entire branch is computed by varying the scalar field's value at the origin $\psi_0$.

We present data sets for rotation numbers $l \in [0, 6]$. These families are characterized and studied using global parameters such as Komar masses and angular momenta. The Komar quantities are chosen instead of the ADM alternatives because of their exponential convergence, which makes them ideal for a non-compactified grid where spatial infinity is an approximation at $r_{\text{bdy}}$. Analyzing the results at fixed rotation integer we determine maximum masses and minimum frequencies. In particular, $M_{\text{max}}^{l = 0} = 0.633\,m_P^2/m$; $M_{\text{max}}^{l = 1} = 1.316\,m_P^2/m$; $M_{\text{max}}^{l = 2} = 2.216\,m_P^2/m$. These latter masses coincide with previous results in literature found by Yoshida and Eriguchi \cite{Yoshida_1997}, and Grandclement, Somé and Gourgoulhon \cite{Grandclement_2014}. Most notably, we also obtain maximum masses and turning-point frequencies for $l > 4$ which are new results in the literature to our knowledge. Given the fact that as the scalar field's amplitude increases, it also approaches the rotation axis, high resolutions are needed to properly resolve and study the field. The new maximum masses for $l > 2$ directly correspond to these limit cases, and their proper study is revealed to be only possible by adding the sixth regularization variable, thus proving again the necessity of adequate regularization in curvilinear formulations at high resolution. 

Future work will concentrate in studying these datasets (including the new high-amplitude regularized members) regarding their stability properties both under 3D and 2D axisymmetric perturbations. Also, given the high versatility of Global Newton Methods, an extension to neighboring problems such as interacting scalar fields and scalar hair is also possible.

%%%%%%%%%%%%%%%%%%%%%%%%%%%
%%%   ACKNOWLEDGMENTS   %%%
%%%%%%%%%%%%%%%%%%%%%%%%%%%

\begin{acknowledgments}
This work was partially supported by CONACyT Network Projects
No. 376127 ``Sombras, lentes y ondas gravitatorias generadas por objetos
compactos astrof\'isicos", and No. 304001 ``Estudio de campos escalares con
aplicaciones en cosmolog\'ia y astrof\'isica", 
and by a CONACyT National Graduate Grant.
\end{acknowledgments}

%%%%%%%%%%%%%%%%%%%%%%
%%%   APPENDIXES   %%%
%%%%%%%%%%%%%%%%%%%%%%

\appendix

\section*{Appendixes}

\section{Complete Regularization Equation}\label{sec:Appendix 1}

Here we present the full equation for the regularization variable $\lambda$ that was obtained following the discussion in Section \ref{sec:Regularization}:
\begin{align}\label{eq:f_lambda}
\begin{split}
    f_\lambda \equiv&\, \partial_\rho^2\lambda + \partial_z^2\lambda + \frac{3}{\rho}\,\partial_\rho \lambda\\
    &\, +  \frac{1}{2H}\,\left(\partial_z H \,\partial_z \lambda - \partial_\rho H\, \partial_\rho \lambda\right) - \frac{2}{A}\,\partial \lambda \cdot \partial H - \frac{\rho^2}{A}\,\partial\lambda\cdot \partial \lambda + \frac{1}{\alpha}\,\left(\partial_z \alpha\, \partial \lambda - \partial_\rho \alpha \, \partial_\rho \lambda\right)\\
    &\, - \frac{4\lambda^2}{A} - \frac{4\lambda \,\rho}{A}\,\partial_\rho \lambda - \frac{2\lambda}{\alpha}\,\left(\frac{\partial_\rho \alpha}{\rho}\right) + \left(\frac{\partial_\rho H}{\rho}\right)\,\left(-\frac{4}{A} + \frac{1}{H}\right)\,\lambda - \left(\frac{\partial_\rho H}{\rho} \right)^2\,\left(\frac{1}{A} + \frac{1}{2H}\right)\\
    &\, + \frac{(\partial_z H)^2}{A\,H}\,\lambda + \frac{(\partial_\rho H)^2}{2H^2}\,\lambda - \frac{2}{\alpha}\,\left(\frac{\partial_\rho H}{\rho}\right)\,\left(\frac{\partial_\rho \alpha}{\rho}\right) - \frac{H^2}{\alpha^2}\,\partial\Omega \cdot \partial \Omega - \frac{A\,H}{\alpha^2}\,(\partial_\rho \Omega)^2\\
    &\, + \frac{2\lambda}{\alpha}\,\partial_\rho^2\alpha + \frac{\lambda}{H}\,\partial_\rho^2H + \left[\frac{1}{\rho}\,\partial_\rho\,\left(\frac{\partial_\rho H}{\rho}\right)\right] + \frac{2H}{\alpha}\,\left[\frac{1}{\rho}\,\partial_\rho\,\left(\frac{\partial_\rho \alpha}{\rho}\right)\right]\\
    &\, + 8\pi\,(\rho^{l-1})^2\,A\,\left(\rho^2\,m^2\,\lambda + 2\,\left(\frac{\partial_\rho \psi}{\rho}\right)\,\left( 2l\,\psi + \rho\,\partial_\rho \psi\right)\right) = 0\,.
\end{split}
\end{align}
Notice how all terms are written in a regularized manner. For example, parentheses explicitly contain the overall-even terms $(1/\rho)\,\partial_\rho u$, and the combination
\begin{equation}
    \frac{1}{\rho^2}\,\left(\partial_\rho^2 - \frac{\partial_\rho u}{\rho}\right) = \left[\frac{1}{\rho}\,\partial_\rho\,\left(\frac{\partial_\rho u}{\rho}\right)\right]\,,
\end{equation}
is also regular by differentiating the even function $((1/\rho)\,\partial_\rho u)$ and then dividing by $\rho$: thus resulting in an even and regular expansion.

\section{NLEQ-ERR Algorithm}\label{sec:Appendix 2}

The Global Newton Method, NLEQ-ERR, is taken from Defulhard \cite{Deuflhard_2011}.

% Spacing is necessary for preprint format.
{\setstretch{1.0}

\begin{algorithm}[H]
\SetAlgoLined
%\KwResult{Write here the result}
 \textbf{Inputs}: Error accuracy $\epsilon$ sufficiently above machine precision, initial iterate $x^0$, right-hand side $F(x^0)$, and damping factor $s_0$.
 
 \textbf{Output}: Newton solution $x^*$ for a convergent algorithm.

    \For{$k = 0,\,1,\,\ldots$}{
    \textbf{1. Step $k$:} Evaluate Jacobian matrix $F'(x^k)$. Solve linear system
    
    $$F'(x^k) \,\Delta x^k = -F(x^k)\,.$$
    
    \textbf{Convergence test}: \If{$\left\Vert \Delta x^k \right\Vert < \epsilon$}{
        \textbf{Stop}. Solution found $x^\star = x^k + \Delta x^k$.}
         \For{$k > 0$}{Compute a prediction value for the damping factor
        $$\mu_k = \frac{\left\Vert \Delta x^{k- 1}\right\Vert \cdot \left\Vert \overline{\Delta x^k}\right\Vert}{\left\Vert \overline{\Delta x^k} - \Delta x^k\right\Vert \cdot \left\Vert \Delta x^k\right\Vert}\,\cdot\,s_{k - 1} \,,\quad s_k = \min\left(1,\,\mu_k\right)\,.$$}

    \textbf{Regularity test}: \If{$s_k < s_{\min}$}{\textbf{Stop}. Convergence failure.}
    
    \textbf{2.} Compute the trial iterate $x^{k+1} = x^k + s_k\,\Delta x^k$ and evaluate $F(x^{k+1})$. Solve the linear system (``old'' Jacobian, ``new'' right-hand side):
    
    $$F'(x^k)\,\overline{\Delta x^{k+1}} = -F(x^{k+1})\,.$$
    
    \textbf{3. } Compute the monitoring quantities 
    
    $$\Theta_k = \frac{\left\Vert \overline{\Delta x^{k+1}}\right\Vert}{\left\Vert \Delta x^k\right\Vert}\,,\quad \mu'_k = \frac{\tfrac{1}{2}\,\left\Vert \Delta x^k \right\Vert \cdot s_k^2}{\left\Vert \overline{\Delta x^{k + 1}} - (1 - s_k)\,\Delta x^k \right\Vert}\,.$$
    
    \eIf{$\Theta_k \geq 1$}{Replace $s_k$ by $s_k' = \min\left(\mu'_k,\,\tfrac{1}{2}\,s_k\right)$. Go to \textbf{Regularity test}.}{Let $s_k' = \min(1,\,\mu_k')$.}
    \uIf{$s_k' = s_k = 1$}{\uIf{$\left\Vert \overline{\Delta x^{k+1}} \right\Vert < \epsilon$}{ \textbf{Stop.} Solution found $u^* = u^{k+1} + \overline{\Delta x^{k+1}}$}}
    \uElseIf{$s_k' > 4\,s_k$}{Replace $s_k$ by $s_k'$ and go to \textbf{2}.}
    \Else{Accept $x^{k+1}$ as new iterate. Continue loop with $k = k + 1$}
    
    }
 \caption{NLEQ-ERR}
\end{algorithm}
}

The algorithm as written in our numerical code \textit{ROTBOSON} has additional features to avoid getting stuck within infinite non-convergent iterations and also uses a local solver, QNERR \cite{Deuflhard_2011}, once the current iteration is determined to be sufficiently close to the final solution, as was detailed in Section \ref{sec:Global Newton Methods}.

%%%%%%%%%%%%%%%%%%%%%%
%%%   REFERENCES   %%%
%%%%%%%%%%%%%%%%%%%%%%

\bibliographystyle{apsrev4-2}
\bibliography{rbs}

\begin{thebibliography}{57}%
\makeatletter
\providecommand \@ifxundefined [1]{%
 \@ifx{#1\undefined}
}%
\providecommand \@ifnum [1]{%
 \ifnum #1\expandafter \@firstoftwo
 \else \expandafter \@secondoftwo
 \fi
}%
\providecommand \@ifx [1]{%
 \ifx #1\expandafter \@firstoftwo
 \else \expandafter \@secondoftwo
 \fi
}%
\providecommand \natexlab [1]{#1}%
\providecommand \enquote  [1]{``#1''}%
\providecommand \bibnamefont  [1]{#1}%
\providecommand \bibfnamefont [1]{#1}%
\providecommand \citenamefont [1]{#1}%
\providecommand \href@noop [0]{\@secondoftwo}%
\providecommand \href [0]{\begingroup \@sanitize@url \@href}%
\providecommand \@href[1]{\@@startlink{#1}\@@href}%
\providecommand \@@href[1]{\endgroup#1\@@endlink}%
\providecommand \@sanitize@url [0]{\catcode `\\12\catcode `\$12\catcode
  `\&12\catcode `\#12\catcode `\^12\catcode `\_12\catcode `\%12\relax}%
\providecommand \@@startlink[1]{}%
\providecommand \@@endlink[0]{}%
\providecommand \url  [0]{\begingroup\@sanitize@url \@url }%
\providecommand \@url [1]{\endgroup\@href {#1}{\urlprefix }}%
\providecommand \urlprefix  [0]{URL }%
\providecommand \Eprint [0]{\href }%
\providecommand \doibase [0]{https://doi.org/}%
\providecommand \selectlanguage [0]{\@gobble}%
\providecommand \bibinfo  [0]{\@secondoftwo}%
\providecommand \bibfield  [0]{\@secondoftwo}%
\providecommand \translation [1]{[#1]}%
\providecommand \BibitemOpen [0]{}%
\providecommand \bibitemStop [0]{}%
\providecommand \bibitemNoStop [0]{.\EOS\space}%
\providecommand \EOS [0]{\spacefactor3000\relax}%
\providecommand \BibitemShut  [1]{\csname bibitem#1\endcsname}%
\let\auto@bib@innerbib\@empty
%</preamble>
\bibitem [{\citenamefont {Yoshida}\ and\ \citenamefont
  {Eriguchi}(1997)}]{Yoshida_1997}%
  \BibitemOpen
  \bibfield  {author} {\bibinfo {author} {\bibfnamefont {S.}~\bibnamefont
  {Yoshida}}\ and\ \bibinfo {author} {\bibfnamefont {Y.}~\bibnamefont
  {Eriguchi}},\ }\href {https://link.aps.org/doi/10.1103/PhysRevD.56.762}
  {\bibfield  {journal} {\bibinfo  {journal} {Phys. Rev. D}\ } (\bibinfo {year}
  {1997})}\BibitemShut {NoStop}%
\bibitem [{\citenamefont {Lai}(2004)}]{Lai_2004}%
  \BibitemOpen
  \bibfield  {author} {\bibinfo {author} {\bibfnamefont {C.}~\bibnamefont
  {Lai}},\ }\emph {\bibinfo {title} {A numerical study of boson stars}},\ \href
  {https://arxiv.org/abs/gr-qc/0410040} {Ph.D. thesis},\ \bibinfo  {school}
  {The University of British Columbia} (\bibinfo {year} {2004})\BibitemShut
  {NoStop}%
\bibitem [{\citenamefont {Grandcl{\'{e}}ment}\ \emph
  {et~al.}(2014)\citenamefont {Grandcl{\'{e}}ment}, \citenamefont
  {Som{\'{e}}},\ and\ \citenamefont {Gourgoulhon}}]{Grandclement_2014}%
  \BibitemOpen
  \bibfield  {author} {\bibinfo {author} {\bibfnamefont {P.}~\bibnamefont
  {Grandcl{\'{e}}ment}}, \bibinfo {author} {\bibfnamefont {C.}~\bibnamefont
  {Som{\'{e}}}},\ and\ \bibinfo {author} {\bibfnamefont {E.}~\bibnamefont
  {Gourgoulhon}},\ }\href {https://doi.org/10.1103/physrevd.90.024068}
  {\bibfield  {journal} {\bibinfo  {journal} {Physical Review D}\ } (\bibinfo
  {year} {2014})}\BibitemShut {NoStop}%
\bibitem [{\citenamefont {Gourgoulhon}(2010)}]{Gourgoulhon_2010}%
  \BibitemOpen
  \bibfield  {author} {\bibinfo {author} {\bibfnamefont {E.}~\bibnamefont
  {Gourgoulhon}},\ }\href@noop {} {\bibinfo {title} {An introduction to the
  theory of rotating relativistic stars}} (\bibinfo {year} {2010}),\ \Eprint
  {https://arxiv.org/abs/arXiv:1003.5015} {arXiv:1003.5015} \BibitemShut
  {NoStop}%
\bibitem [{\citenamefont {Alcubierre}(2012)}]{Alcubierre_2012}%
  \BibitemOpen
  \bibfield  {author} {\bibinfo {author} {\bibfnamefont {M.}~\bibnamefont
  {Alcubierre}},\ }\href
  {https://oxford.universitypressscholarship.com/view/10.1093/acprof:oso/9780199205677.001.0001/acprof-9780199205677}
  {\emph {\bibinfo {title} {Introduction to 3 + 1 numerical relativity}}}\
  (\bibinfo  {publisher} {Oxford University Press},\ \bibinfo {year}
  {2012})\BibitemShut {NoStop}%
\bibitem [{\citenamefont {Ruiz}\ \emph {et~al.}(2008)\citenamefont {Ruiz},
  \citenamefont {Alcubierre},\ and\ \citenamefont
  {N{\'u}{\~{n}}ez}}]{Ruiz_2008}%
  \BibitemOpen
  \bibfield  {author} {\bibinfo {author} {\bibfnamefont {M.}~\bibnamefont
  {Ruiz}}, \bibinfo {author} {\bibfnamefont {M.}~\bibnamefont {Alcubierre}},\
  and\ \bibinfo {author} {\bibfnamefont {D.}~\bibnamefont {N{\'u}{\~{n}}ez}},\
  }\href {https://doi.org/10.1007/s10714-007-0522-3} {\bibfield  {journal}
  {\bibinfo  {journal} {General Relativity and Gravitation}\ } (\bibinfo {year}
  {2008})}\BibitemShut {NoStop}%
\bibitem [{\citenamefont {Torres}(2012)}]{Torres_2012}%
  \BibitemOpen
  \bibfield  {author} {\bibinfo {author} {\bibfnamefont {J.~M.}\ \bibnamefont
  {Torres}},\ }\href {https://doi.org/10.1063/1.4748531} {\bibfield  {journal}
  {\bibinfo  {journal} {AIP Conference Proceedings}\ } (\bibinfo {year}
  {2012})}\BibitemShut {NoStop}%
\bibitem [{\citenamefont {Deuflhard}(2011)}]{Deuflhard_2011}%
  \BibitemOpen
  \bibfield  {author} {\bibinfo {author} {\bibfnamefont {P.}~\bibnamefont
  {Deuflhard}},\ }\href {https://doi.org/10.1007/978-3-642-23899-4} {\emph
  {\bibinfo {title} {Newton methods for nonlinear problems affine invariance
  and adaptive algorithms}}}\ (\bibinfo  {publisher} {Springer},\ \bibinfo
  {year} {2011})\BibitemShut {NoStop}%
\bibitem [{\citenamefont {Petra}\ \emph
  {et~al.}(2014{\natexlab{a}})\citenamefont {Petra}, \citenamefont {Schenk},\
  and\ \citenamefont {Anitescu}}]{PARDISO_1}%
  \BibitemOpen
  \bibfield  {author} {\bibinfo {author} {\bibfnamefont {C.~G.}\ \bibnamefont
  {Petra}}, \bibinfo {author} {\bibfnamefont {O.}~\bibnamefont {Schenk}},\ and\
  \bibinfo {author} {\bibfnamefont {M.}~\bibnamefont {Anitescu}},\ }\href
  {https://doi.org/10.1109/MCSE.2014.53} {\bibfield  {journal} {\bibinfo
  {journal} {Computing in Science Engineering}\ } (\bibinfo {year}
  {2014}{\natexlab{a}})}\BibitemShut {NoStop}%
\bibitem [{\citenamefont {Petra}\ \emph
  {et~al.}(2014{\natexlab{b}})\citenamefont {Petra}, \citenamefont {Schenk},
  \citenamefont {Lubin},\ and\ \citenamefont {Gäertner}}]{PARDISO_2}%
  \BibitemOpen
  \bibfield  {author} {\bibinfo {author} {\bibfnamefont {C.~G.}\ \bibnamefont
  {Petra}}, \bibinfo {author} {\bibfnamefont {O.}~\bibnamefont {Schenk}},
  \bibinfo {author} {\bibfnamefont {M.}~\bibnamefont {Lubin}},\ and\ \bibinfo
  {author} {\bibfnamefont {K.}~\bibnamefont {Gäertner}},\ }\href
  {https://doi.org/10.1137/130908737} {\bibfield  {journal} {\bibinfo
  {journal} {SIAM Journal on Scientific Computing}\ } (\bibinfo {year}
  {2014}{\natexlab{b}})}\BibitemShut {NoStop}%
\bibitem [{\citenamefont {Ontanon}(2021{\natexlab{a}})}]{ROTBOSON}%
  \BibitemOpen
  \bibfield  {author} {\bibinfo {author} {\bibfnamefont {S.}~\bibnamefont
  {Ontanon}},\ }\href@noop {} {\bibinfo {title} {{ROTBOSON}: Rotating boson
  stars initial data for numerical relativity}},\ \bibinfo {howpublished}
  {\url{https://github.com/sontanon/ROTBOSON}} (\bibinfo {year}
  {2021}{\natexlab{a}})\BibitemShut {NoStop}%
\bibitem [{\citenamefont {Liebling}\ and\ \citenamefont
  {Palenzuela}(2017)}]{Liebling_2017}%
  \BibitemOpen
  \bibfield  {author} {\bibinfo {author} {\bibfnamefont {S.~L.}\ \bibnamefont
  {Liebling}}\ and\ \bibinfo {author} {\bibfnamefont {C.}~\bibnamefont
  {Palenzuela}},\ }\href {https://doi.org/10.1007/s41114-017-0007-y} {\bibfield
   {journal} {\bibinfo  {journal} {Living Reviews in Relativity}\ } (\bibinfo
  {year} {2017})}\BibitemShut {NoStop}%
\bibitem [{\citenamefont {Kaup}(1968)}]{Kaup_1968}%
  \BibitemOpen
  \bibfield  {author} {\bibinfo {author} {\bibfnamefont {D.~J.}\ \bibnamefont
  {Kaup}},\ }\href {https://link.aps.org/doi/10.1103/PhysRev.172.1331}
  {\bibfield  {journal} {\bibinfo  {journal} {Phys. Rev.}\ } (\bibinfo {year}
  {1968})}\BibitemShut {NoStop}%
\bibitem [{\citenamefont {Ruffini}\ and\ \citenamefont
  {Bonzzola}(1969)}]{Ruffini_1969}%
  \BibitemOpen
  \bibfield  {author} {\bibinfo {author} {\bibfnamefont {R.}~\bibnamefont
  {Ruffini}}\ and\ \bibinfo {author} {\bibfnamefont {S.}~\bibnamefont
  {Bonzzola}},\ }\href {https://link.aps.org/doi/10.1103/PhysRev.187.1767}
  {\bibfield  {journal} {\bibinfo  {journal} {Phys. Rev.}\ } (\bibinfo {year}
  {1969})}\BibitemShut {NoStop}%
\bibitem [{\citenamefont {{Einstein}}(1915)}]{Einstein_1915}%
  \BibitemOpen
  \bibfield  {author} {\bibinfo {author} {\bibfnamefont {A.}~\bibnamefont
  {{Einstein}}},\ }\href {http://adsabs.harvard.edu/abs/1915SPAW.......844E}
  {\bibfield  {journal} {\bibinfo  {journal} {Sitzungsberichte der
  K{\"o}niglich Preu{\ss}ischen Akademie der Wissenschaften (Berlin), Seite
  844-847.}\ } (\bibinfo {year} {1915})}\BibitemShut {NoStop}%
\bibitem [{\citenamefont {Wheeler}(1955)}]{Wheeler_1955}%
  \BibitemOpen
  \bibfield  {author} {\bibinfo {author} {\bibfnamefont {J.~A.}\ \bibnamefont
  {Wheeler}},\ }\href {https://link.aps.org/doi/10.1103/PhysRev.97.511}
  {\bibfield  {journal} {\bibinfo  {journal} {Phys. Rev.}\ } (\bibinfo {year}
  {1955})}\BibitemShut {NoStop}%
\bibitem [{\citenamefont {Derrick}(1964)}]{Derrick_1964}%
  \BibitemOpen
  \bibfield  {author} {\bibinfo {author} {\bibfnamefont {G.~H.}\ \bibnamefont
  {Derrick}},\ }\href {https://doi.org/10.1063/1.1704233} {\bibfield  {journal}
  {\bibinfo  {journal} {Journal of Mathematical Physics}\ } (\bibinfo {year}
  {1964})}\BibitemShut {NoStop}%
\bibitem [{\citenamefont {Diez-Tejedor}\ and\ \citenamefont
  {Gonzalez-Morales}(2013)}]{Diez-Tejedor_2013}%
  \BibitemOpen
  \bibfield  {author} {\bibinfo {author} {\bibfnamefont {A.}~\bibnamefont
  {Diez-Tejedor}}\ and\ \bibinfo {author} {\bibfnamefont {A.~X.}\ \bibnamefont
  {Gonzalez-Morales}},\ }\href
  {https://link.aps.org/doi/10.1103/PhysRevD.88.067302} {\bibfield  {journal}
  {\bibinfo  {journal} {Phys. Rev. D}\ } (\bibinfo {year} {2013})}\BibitemShut
  {NoStop}%
\bibitem [{\citenamefont {Schunck}\ and\ \citenamefont
  {Mielke}(2003)}]{Schunck_2003}%
  \BibitemOpen
  \bibfield  {author} {\bibinfo {author} {\bibfnamefont {F.~E.}\ \bibnamefont
  {Schunck}}\ and\ \bibinfo {author} {\bibfnamefont {E.~W.}\ \bibnamefont
  {Mielke}},\ }\href {https://doi.org/10.1088/0264-9381/20/20/201} {\bibfield
  {journal} {\bibinfo  {journal} {Classical and Quantum Gravity}\ } (\bibinfo
  {year} {2003})}\BibitemShut {NoStop}%
\bibitem [{\citenamefont {Ureña-López}\ and\ \citenamefont
  {Bernal}(2010)}]{Urena-Lopez_2010}%
  \BibitemOpen
  \bibfield  {author} {\bibinfo {author} {\bibfnamefont {L.~A.}\ \bibnamefont
  {Ureña-López}}\ and\ \bibinfo {author} {\bibfnamefont {A.}~\bibnamefont
  {Bernal}},\ }\href {https://link.aps.org/doi/10.1103/PhysRevD.82.123535}
  {\bibfield  {journal} {\bibinfo  {journal} {Phys. Rev. D}\ } (\bibinfo {year}
  {2010})}\BibitemShut {NoStop}%
\bibitem [{\citenamefont {Guzm\'an}\ and\ \citenamefont
  {Rueda-Becerril}(2009)}]{Guzman_2009}%
  \BibitemOpen
  \bibfield  {author} {\bibinfo {author} {\bibfnamefont {F.~S.}\ \bibnamefont
  {Guzm\'an}}\ and\ \bibinfo {author} {\bibfnamefont {J.~M.}\ \bibnamefont
  {Rueda-Becerril}},\ }\href
  {https://link.aps.org/doi/10.1103/PhysRevD.80.084023} {\bibfield  {journal}
  {\bibinfo  {journal} {Phys. Rev. D}\ } (\bibinfo {year} {2009})}\BibitemShut
  {NoStop}%
\bibitem [{\citenamefont {Barranco}\ and\ \citenamefont
  {Bernal}(2011)}]{Barranco_2011}%
  \BibitemOpen
  \bibfield  {author} {\bibinfo {author} {\bibfnamefont {J.}~\bibnamefont
  {Barranco}}\ and\ \bibinfo {author} {\bibfnamefont {A.}~\bibnamefont
  {Bernal}},\ }\href {https://aip.scitation.org/doi/abs/10.1063/1.3647542}
  {\bibfield  {journal} {\bibinfo  {journal} {AIP Conference Proceedings}\ }
  (\bibinfo {year} {2011})}\BibitemShut {NoStop}%
\bibitem [{\citenamefont {Colpi}\ \emph {et~al.}(1986)\citenamefont {Colpi},
  \citenamefont {Shapiro},\ and\ \citenamefont {Wasserman}}]{Colpi_1986}%
  \BibitemOpen
  \bibfield  {author} {\bibinfo {author} {\bibfnamefont {M.}~\bibnamefont
  {Colpi}}, \bibinfo {author} {\bibfnamefont {S.~L.}\ \bibnamefont {Shapiro}},\
  and\ \bibinfo {author} {\bibfnamefont {I.}~\bibnamefont {Wasserman}},\ }\href
  {https://link.aps.org/doi/10.1103/PhysRevLett.57.2485} {\bibfield  {journal}
  {\bibinfo  {journal} {Phys. Rev. Lett.}\ } (\bibinfo {year}
  {1986})}\BibitemShut {NoStop}%
\bibitem [{\citenamefont {Friedberg}\ \emph {et~al.}(1987)\citenamefont
  {Friedberg}, \citenamefont {Lee},\ and\ \citenamefont
  {Pang}}]{Friedberg_1987}%
  \BibitemOpen
  \bibfield  {author} {\bibinfo {author} {\bibfnamefont {R.}~\bibnamefont
  {Friedberg}}, \bibinfo {author} {\bibfnamefont {T.~D.}\ \bibnamefont {Lee}},\
  and\ \bibinfo {author} {\bibfnamefont {Y.}~\bibnamefont {Pang}},\ }\href
  {https://link.aps.org/doi/10.1103/PhysRevD.35.3640} {\bibfield  {journal}
  {\bibinfo  {journal} {Phys. Rev. D}\ } (\bibinfo {year} {1987})}\BibitemShut
  {NoStop}%
\bibitem [{\citenamefont {Alcubierre}\ \emph {et~al.}(2018)\citenamefont
  {Alcubierre}, \citenamefont {Barranco}, \citenamefont {Bernal}, \citenamefont
  {Degollado}, \citenamefont {Diez-Tejedor}, \citenamefont {Megevand},
  \citenamefont {N{\'{u}}{\~{n}}ez},\ and\ \citenamefont
  {Sarbach}}]{Alcubierre_2018}%
  \BibitemOpen
  \bibfield  {author} {\bibinfo {author} {\bibfnamefont {M.}~\bibnamefont
  {Alcubierre}}, \bibinfo {author} {\bibfnamefont {J.}~\bibnamefont
  {Barranco}}, \bibinfo {author} {\bibfnamefont {A.}~\bibnamefont {Bernal}},
  \bibinfo {author} {\bibfnamefont {J.~C.}\ \bibnamefont {Degollado}}, \bibinfo
  {author} {\bibfnamefont {A.}~\bibnamefont {Diez-Tejedor}}, \bibinfo {author}
  {\bibfnamefont {M.}~\bibnamefont {Megevand}}, \bibinfo {author}
  {\bibfnamefont {D.}~\bibnamefont {N{\'{u}}{\~{n}}ez}},\ and\ \bibinfo
  {author} {\bibfnamefont {O.}~\bibnamefont {Sarbach}},\ }\href
  {https://doi.org/10.1088/1361-6382/aadcb6} {\bibfield  {journal} {\bibinfo
  {journal} {Classical and Quantum Gravity}\ } (\bibinfo {year}
  {2018})}\BibitemShut {NoStop}%
\bibitem [{\citenamefont {Silveira}\ and\ \citenamefont
  {de~Sousa}(1995)}]{Silveira_1995}%
  \BibitemOpen
  \bibfield  {author} {\bibinfo {author} {\bibfnamefont {V.}~\bibnamefont
  {Silveira}}\ and\ \bibinfo {author} {\bibfnamefont {C.~M.~G.}\ \bibnamefont
  {de~Sousa}},\ }\href {https://link.aps.org/doi/10.1103/PhysRevD.52.5724}
  {\bibfield  {journal} {\bibinfo  {journal} {Phys. Rev. D}\ } (\bibinfo {year}
  {1995})}\BibitemShut {NoStop}%
\bibitem [{\citenamefont {Mielke}(2016)}]{Mielke_2016}%
  \BibitemOpen
  \bibfield  {author} {\bibinfo {author} {\bibfnamefont {E.~W.}\ \bibnamefont
  {Mielke}},\ }\bibinfo {title} {Rotating boson stars},\ in\ \href
  {https://doi.org/10.1007/978-3-319-31299-6_6} {\emph {\bibinfo {booktitle}
  {At the Frontier of Spacetime: Scalar-Tensor Theory, Bells Inequality, Machs
  Principle, Exotic Smoothness}}},\ \bibinfo {editor} {edited by\ \bibinfo
  {editor} {\bibfnamefont {T.}~\bibnamefont {Asselmeyer-Maluga}}}\ (\bibinfo
  {publisher} {Springer International Publishing},\ \bibinfo {year} {2016})\
  pp.\ \bibinfo {pages} {115--131}\BibitemShut {NoStop}%
\bibitem [{\citenamefont {Smarr}\ and\ \citenamefont {York}(1978)}]{York_1978}%
  \BibitemOpen
  \bibfield  {author} {\bibinfo {author} {\bibfnamefont {L.}~\bibnamefont
  {Smarr}}\ and\ \bibinfo {author} {\bibfnamefont {J.~W.}\ \bibnamefont
  {York}},\ }\href {https://doi.org/10.1103/PhysRevD.17.2529} {\bibfield
  {journal} {\bibinfo  {journal} {Phys. Rev. D}\ } (\bibinfo {year}
  {1978})}\BibitemShut {NoStop}%
\bibitem [{\citenamefont {Lewis}(1932)}]{Lewis_1932}%
  \BibitemOpen
  \bibfield  {author} {\bibinfo {author} {\bibfnamefont {T.}~\bibnamefont
  {Lewis}},\ }\href
  {https://royalsocietypublishing.org/doi/10.1098/rspa.1932.0073} {\bibfield
  {journal} {\bibinfo  {journal} {Proceedings of the Royal Society A}\ }
  (\bibinfo {year} {1932})}\BibitemShut {NoStop}%
\bibitem [{\citenamefont {Papapetrou}(1945)}]{Papapetrou_1945}%
  \BibitemOpen
  \bibfield  {author} {\bibinfo {author} {\bibfnamefont {A.}~\bibnamefont
  {Papapetrou}},\ }\href {https://www.jstor.org/stable/20488481} {\bibfield
  {journal} {\bibinfo  {journal} {Proceedings of the Roya Irish Academy.
  Section A: Mathematical and Physical Sciences}\ } (\bibinfo {year}
  {1945})}\BibitemShut {NoStop}%
\bibitem [{\citenamefont {Gustafson}(1998)}]{Gustafson_1998}%
  \BibitemOpen
  \bibfield  {author} {\bibinfo {author} {\bibfnamefont {K.}~\bibnamefont
  {Gustafson}},\ }\href {http://dx.doi.org/10.1090/conm/218} {\bibfield
  {journal} {\bibinfo  {journal} {Contemporary Mathematics}\ } (\bibinfo {year}
  {1998})}\BibitemShut {NoStop}%
\bibitem [{\citenamefont {L{\"u}then}\ \emph {et~al.}(2018)\citenamefont
  {L{\"u}then}, \citenamefont {Juntunen},\ and\ \citenamefont
  {Stenberg}}]{Nora_2018}%
  \BibitemOpen
  \bibfield  {author} {\bibinfo {author} {\bibfnamefont {N.}~\bibnamefont
  {L{\"u}then}}, \bibinfo {author} {\bibfnamefont {M.}~\bibnamefont
  {Juntunen}},\ and\ \bibinfo {author} {\bibfnamefont {R.}~\bibnamefont
  {Stenberg}},\ }\href {https://doi.org/10.1007/s00211-017-0927-1} {\bibfield
  {journal} {\bibinfo  {journal} {Numerische Mathematik}\ } (\bibinfo {year}
  {2018})}\BibitemShut {NoStop}%
\bibitem [{\citenamefont {Arfken}\ and\ \citenamefont
  {Weber}(2005)}]{Arfken_2005}%
  \BibitemOpen
  \bibfield  {author} {\bibinfo {author} {\bibfnamefont {G.}~\bibnamefont
  {Arfken}}\ and\ \bibinfo {author} {\bibfnamefont {H.}~\bibnamefont {Weber}},\
  }\href {https://doi.org/10.1016/C2009-0-30629-7} {\emph {\bibinfo {title}
  {Mathematical Methods for Physicists}}},\ Mathematical Methods for
  Physicists\ (\bibinfo  {publisher} {Elsevier},\ \bibinfo {year}
  {2005})\BibitemShut {NoStop}%
\bibitem [{\citenamefont {Wald}(1984)}]{Wald_1984}%
  \BibitemOpen
  \bibfield  {author} {\bibinfo {author} {\bibfnamefont {R.~M.}\ \bibnamefont
  {Wald}},\ }\href {https://cds.cern.ch/record/106274} {\emph {\bibinfo {title}
  {General relativity}}}\ (\bibinfo  {publisher} {Chicago Univ. Press},\
  \bibinfo {address} {Chicago, IL},\ \bibinfo {year} {1984})\BibitemShut
  {NoStop}%
\bibitem [{\citenamefont {Alcubierre}\ \emph {et~al.}(2000)\citenamefont
  {Alcubierre}, \citenamefont {Brandt}, \citenamefont {Brügmann},
  \citenamefont {Gundlach}, \citenamefont {Massó}, \citenamefont {Seidel},\
  and\ \citenamefont {Walker}}]{Alcubierre_2000}%
  \BibitemOpen
  \bibfield  {author} {\bibinfo {author} {\bibfnamefont {M.}~\bibnamefont
  {Alcubierre}}, \bibinfo {author} {\bibfnamefont {S.}~\bibnamefont {Brandt}},
  \bibinfo {author} {\bibfnamefont {B.}~\bibnamefont {Brügmann}}, \bibinfo
  {author} {\bibfnamefont {C.}~\bibnamefont {Gundlach}}, \bibinfo {author}
  {\bibfnamefont {J.}~\bibnamefont {Massó}}, \bibinfo {author} {\bibfnamefont
  {E.}~\bibnamefont {Seidel}},\ and\ \bibinfo {author} {\bibfnamefont
  {P.}~\bibnamefont {Walker}},\ }\href
  {https://doi.org/10.1088/0264-9381/17/11/301} {\bibfield  {journal} {\bibinfo
   {journal} {Classical and Quantum Gravity}\ } (\bibinfo {year}
  {2000})}\BibitemShut {NoStop}%
\bibitem [{\citenamefont {Kollerstrom}(1992)}]{Kollerstrom_1992}%
  \BibitemOpen
  \bibfield  {author} {\bibinfo {author} {\bibfnamefont {N.}~\bibnamefont
  {Kollerstrom}},\ }\href {https://doi.org/10.1017/s0007087400029150}
  {\bibfield  {journal} {\bibinfo  {journal} {The British Journal for the
  History of Science}\ } (\bibinfo {year} {1992})}\BibitemShut {NoStop}%
\bibitem [{\citenamefont {Ypma}(1984)}]{Ypma_1984}%
  \BibitemOpen
  \bibfield  {author} {\bibinfo {author} {\bibfnamefont {T.~J.}\ \bibnamefont
  {Ypma}},\ }\href {https://doi.org/10.1137/0721040} {\bibfield  {journal}
  {\bibinfo  {journal} {SIAM Journal on Numerical Analysis}\ } (\bibinfo {year}
  {1984})}\BibitemShut {NoStop}%
\bibitem [{\citenamefont {Bank}\ and\ \citenamefont {Rose}(1981)}]{Bank_1981}%
  \BibitemOpen
  \bibfield  {author} {\bibinfo {author} {\bibfnamefont {R.~E.}\ \bibnamefont
  {Bank}}\ and\ \bibinfo {author} {\bibfnamefont {D.~J.}\ \bibnamefont
  {Rose}},\ }\href {https://doi.org/10.1007/bf01398257} {\bibfield  {journal}
  {\bibinfo  {journal} {Numerische Mathematik}\ } (\bibinfo {year}
  {1981})}\BibitemShut {NoStop}%
\bibitem [{\citenamefont {LeVeque}(2007)}]{Leveque_2007}%
  \BibitemOpen
  \bibfield  {author} {\bibinfo {author} {\bibfnamefont {R.}~\bibnamefont
  {LeVeque}},\ }\href {https://doi.org/10.1137/1.9780898717839} {\emph
  {\bibinfo {title} {Finite Difference Methods for Ordinary and Partial
  Differential Equations: Steady-State and Time-Dependent Problems}}},\ Other
  Titles in Applied Mathematics\ (\bibinfo  {publisher} {Society for Industrial
  and Applied Mathematics},\ \bibinfo {year} {2007})\BibitemShut {NoStop}%
\bibitem [{\citenamefont {Kantorovich}\ and\ \citenamefont
  {Akilov}(1982)}]{Kantorovich_1982}%
  \BibitemOpen
  \bibfield  {author} {\bibinfo {author} {\bibfnamefont {L.~V.}\ \bibnamefont
  {Kantorovich}}\ and\ \bibinfo {author} {\bibfnamefont {G.~P.}\ \bibnamefont
  {Akilov}},\ }\href {https://doi.org/10.1016/C2013-0-03044-7} {\emph {\bibinfo
  {title} {Functional analysis}}}\ (\bibinfo  {publisher} {Pergamon},\ \bibinfo
  {year} {1982})\BibitemShut {NoStop}%
\bibitem [{\citenamefont {Ortega}\ and\ \citenamefont
  {Rheinboldt}(2000)}]{Ortega_2000}%
  \BibitemOpen
  \bibfield  {author} {\bibinfo {author} {\bibfnamefont {J.~M.}\ \bibnamefont
  {Ortega}}\ and\ \bibinfo {author} {\bibfnamefont {W.~C.}\ \bibnamefont
  {Rheinboldt}},\ }\href {https://doi.org/10.1016/C2013-0-11263-9} {\emph
  {\bibinfo {title} {Iterative solution of nonlinear equations in several
  variables}}}\ (\bibinfo  {publisher} {Society for Industrial and Applied
  Mathematics},\ \bibinfo {year} {2000})\BibitemShut {NoStop}%
\bibitem [{\citenamefont {Deuflhard}(1975)}]{Deuflhard_1975}%
  \BibitemOpen
  \bibfield  {author} {\bibinfo {author} {\bibfnamefont {P.}~\bibnamefont
  {Deuflhard}},\ }\href {https://doi.org/10.1007/bfb0079167} {\bibfield
  {journal} {\bibinfo  {journal} {Lecture Notes in Mathematics Optimization and
  Optimal Control}\ } (\bibinfo {year} {1975})}\BibitemShut {NoStop}%
\bibitem [{\citenamefont {Cauchy}(1847)}]{Cauchy_1847}%
  \BibitemOpen
  \bibfield  {author} {\bibinfo {author} {\bibfnamefont {A.-L.}\ \bibnamefont
  {Cauchy}},\ }\href {https://doi.org/10.1017/cbo9780511702396.063} {\bibfield
  {journal} {\bibinfo  {journal} {Oeuvres complètes}\ } (\bibinfo {year}
  {1847})}\BibitemShut {NoStop}%
\bibitem [{\citenamefont {Levenberg}(1944)}]{Levenberg_1944}%
  \BibitemOpen
  \bibfield  {author} {\bibinfo {author} {\bibfnamefont {K.}~\bibnamefont
  {Levenberg}},\ }\href {https://doi.org/10.1090/qam/10666} {\bibfield
  {journal} {\bibinfo  {journal} {Quarterly of Applied Mathematics}\ }
  (\bibinfo {year} {1944})}\BibitemShut {NoStop}%
\bibitem [{\citenamefont {Marquardt}(1963)}]{Marquardt_1963}%
  \BibitemOpen
  \bibfield  {author} {\bibinfo {author} {\bibfnamefont {D.~W.}\ \bibnamefont
  {Marquardt}},\ }\href {http://www.jstor.org/stable/2098941} {\bibfield
  {journal} {\bibinfo  {journal} {Journal of the Society for Industrial and
  Applied Mathematics}\ } (\bibinfo {year} {1963})}\BibitemShut {NoStop}%
\bibitem [{\citenamefont {Nowak}\ and\ \citenamefont {Weimann}(1991)}]{NLEQ1}%
  \BibitemOpen
  \bibfield  {author} {\bibinfo {author} {\bibfnamefont {U.}~\bibnamefont
  {Nowak}}\ and\ \bibinfo {author} {\bibfnamefont {L.}~\bibnamefont
  {Weimann}},\ }\href {https://opus4.kobv.de/opus4-zib/files/485/TR-91-10.pdf}
  {\emph {\bibinfo {title} {A Family of Newton Codes for Systems of Highly
  Nonlinear Equations}}},\ \bibinfo {type} {Tech. Rep.}\ (\bibinfo
  {institution} {Zuse Institute Berlin},\ \bibinfo {year} {1991})\BibitemShut
  {NoStop}%
\bibitem [{\citenamefont {Dagum}\ and\ \citenamefont
  {Menon}(1998)}]{OpenMP_1998}%
  \BibitemOpen
  \bibfield  {author} {\bibinfo {author} {\bibfnamefont {L.}~\bibnamefont
  {Dagum}}\ and\ \bibinfo {author} {\bibfnamefont {R.}~\bibnamefont {Menon}},\
  }\href {https://doi.org/10.1109/99.660313} {\bibfield  {journal} {\bibinfo
  {journal} {IEEE Comput. Sci. Eng.}\ } (\bibinfo {year} {1998})}\BibitemShut
  {NoStop}%
\bibitem [{\citenamefont {Torres}(2016)}]{Torres_2016}%
  \BibitemOpen
  \bibfield  {author} {\bibinfo {author} {\bibfnamefont {J.~M.}\ \bibnamefont
  {Torres}},\ }\emph {\bibinfo {title} {Dinámica de materia cargada en
  relatividad general}},\ \href
  {https://repositorio.unam.mx/contenidos/dinamica-de-materia-cargada-en-relatividad-general-31-84942?c=pQ8wXB&d=false&q=Din%C3%A1mica_._de_._materia_._cargada_._en_._relatividad_._general&i=1&v=1&t=search_0&as=0#}
  {Ph.D. thesis},\ \bibinfo  {school} {Universidad Nacional Autónoma de
  México, Instituto de Ciencias Nucleares} (\bibinfo {year}
  {2016})\BibitemShut {NoStop}%
\bibitem [{\citenamefont {Ontanon}(2018)}]{Ontanon_2018}%
  \BibitemOpen
  \bibfield  {author} {\bibinfo {author} {\bibfnamefont {S.}~\bibnamefont
  {Ontanon}},\ }\href
  {https://repositorio.unam.mx/contenidos/resolvedor-eliptico-para-relatividad-numerica-y-su-aplicacion-en-ondas-de-brill-170349?c=pQ8wXB&d=false&q=resolvedor_._eliptico_._brill&i=1&v=1&t=search_0&as=0#}
  {\bibinfo {title} {Resolvedor elíptico para relatividad numérica y su
  aplicación en ondas de {B}rill}},\ \bibinfo {howpublished} {Bachelor's
  Thesis} (\bibinfo {year} {2018})\BibitemShut {NoStop}%
\bibitem [{\citenamefont {Corporation}(2018)}]{MKL_PARDISO_2018}%
  \BibitemOpen
  \bibfield  {author} {\bibinfo {author} {\bibfnamefont {I.}~\bibnamefont
  {Corporation}},\ }\href@noop {} {\bibinfo {title} {Intel mkl pardiso -
  parallel direct sparse solver interface}} (\bibinfo {year} {2018}),\ \bibinfo
  {note} {online}\BibitemShut {NoStop}%
\bibitem [{\citenamefont {Tewarson}(1973)}]{Tewarson_1973}%
  \BibitemOpen
  \bibinfo {editor} {\bibfnamefont {R.~P.}\ \bibnamefont {Tewarson}},\ ed.,\
  \href
  {https://www.sciencedirect.com/bookseries/mathematics-in-science-and-engineering/vol/99}
  {\emph {\bibinfo {title} {Sparse Matrices}}},\ Mathematics in Science and
  Engineering\ (\bibinfo  {publisher} {Elsevier Science},\ \bibinfo {year}
  {1973})\BibitemShut {NoStop}%
\bibitem [{\citenamefont {Gupta}\ and\ \citenamefont
  {Kumar}(1995)}]{Gupta_1995}%
  \BibitemOpen
  \bibfield  {author} {\bibinfo {author} {\bibfnamefont {A.}~\bibnamefont
  {Gupta}}\ and\ \bibinfo {author} {\bibfnamefont {V.}~\bibnamefont {Kumar}},\
  }in\ \href {https://doi.org/10.1109/SUPERC.1995.242069} {\emph {\bibinfo
  {booktitle} {Proceedings of the IEEE/ACM SC95 Conference}}}\ (\bibinfo {year}
  {1995})\BibitemShut {NoStop}%
\bibitem [{\citenamefont {Inc.}(2020)}]{Mathematica_2018}%
  \BibitemOpen
  \bibfield  {author} {\bibinfo {author} {\bibfnamefont {W.~R.}\ \bibnamefont
  {Inc.}},\ }\href {https://www.wolfram.com/mathematica} {\bibinfo {title}
  {Mathematica, {V}ersion 12.2}} (\bibinfo {year} {2020}),\ \bibinfo {note}
  {champaign, IL, 2020}\BibitemShut {NoStop}%
\bibitem [{\citenamefont {Schenk}\ and\ \citenamefont
  {Gärtner}(2020)}]{Schenk_2018}%
  \BibitemOpen
  \bibfield  {author} {\bibinfo {author} {\bibfnamefont {O.}~\bibnamefont
  {Schenk}}\ and\ \bibinfo {author} {\bibfnamefont {K.}~\bibnamefont
  {Gärtner}},\ }\href {https://pardiso-project.org/manual/manual.pdf} {\emph
  {\bibinfo {title} {PARDISO User Guide}}},\ \bibinfo {organization} {PARDISO
  Project} (\bibinfo {year} {2020})\BibitemShut {NoStop}%
\bibitem [{\citenamefont {Forum}(1994)}]{MPI_1994}%
  \BibitemOpen
  \bibfield  {author} {\bibinfo {author} {\bibfnamefont {M.~P.}\ \bibnamefont
  {Forum}},\ }\href {https://www.mpi-forum.org/docs/mpi-3.0/mpi30-report.pdf}
  {\emph {\bibinfo {title} {MPI: A Message-Passing Interface Standard}}},\
  \bibinfo {type} {Tech. Rep.}\ (\bibinfo  {institution} {University of
  Tennessee},\ \bibinfo {address} {Knoxville, TN, USA},\ \bibinfo {year}
  {1994})\BibitemShut {NoStop}%
\bibitem [{\citenamefont {Richardson}(1911)}]{Richardson_1911}%
  \BibitemOpen
  \bibfield  {author} {\bibinfo {author} {\bibfnamefont {L.~F.}\ \bibnamefont
  {Richardson}},\ }\href {https://doi.org/10.1098/rsta.1911.0009} {\bibfield
  {journal} {\bibinfo  {journal} {Philosophical Transactions of the Royal
  Society of London. Series A, Containing Papers of a Mathematical or Physical
  Character}\ } (\bibinfo {year} {1911})}\BibitemShut {NoStop}%
\bibitem [{\citenamefont {Ontanon}(2021{\natexlab{b}})}]{SPHBOSON}%
  \BibitemOpen
  \bibfield  {author} {\bibinfo {author} {\bibfnamefont {S.}~\bibnamefont
  {Ontanon}},\ }\href@noop {} {\bibinfo {title} {{SPHBOSON}: Spherical boson
  stars initial data for numerical relativity}},\ \bibinfo {howpublished}
  {\url{https://github.com/sontanon/SPHBOSON}} (\bibinfo {year}
  {2021}{\natexlab{b}})\BibitemShut {NoStop}%
\end{thebibliography}%

%%%%%%%%%%%%%%%
%%%   END   %%%
%%%%%%%%%%%%%%%

\end{document}